\documentclass[aps,pre,reprint,superscriptaddress,preprintnumbers,amsmath,amssymb]{revtex4-2}
\usepackage{graphicx}
\usepackage{dcolumn}% Align table columns on decimal point
\usepackage{bm}% bold math
\usepackage{hyperref}% add hypertext capabilities
\usepackage[version=4]{mhchem}
\begin{document}
	
	%\preprint{APS/123-QED}
	
	\title{Memory of rotation in residual stress of paste}% Force line breaks with \\
	
	\author{Hiroki Matsuda}
	%  \altaffiliation[Also at ]{Physics Department, XYZ University.}%Lines break automatically or can be forced with \\
	\author{Michio Otsuki}%
	\email{m.otsuki.es@osaka-u.ac.jp}
	\affiliation{%
		Graduate School of Engineering Science, Osaka University, Toyonaka 560-8531, Japan
	}%
	
	% \collaboration{MUSO Collaboration}%\noaffiliation
	%  \homepage{http://www.Second.institution.edu/~Charlie.Author}
	% \affiliation{
		%  Second institution and/or address\\
		%  This line break forced% with \\
		% }%
	% \affiliation{
		%  Third institution, the second for Charlie Author
	% \affiliation{%
		%  Authors' institution and/or address\\
		%  This line break forced with \textbackslash\textbackslash
		% }%
	
	% \collaboration{CLEO Collaboration}%\noaffiliation
	
	%              %  but any date may be explicitly specified
	
	\begin{abstract}
We numerically investigate the stress distribution in pastes after horizontal rotation by using an elasto-plastic model.
	    Residual stress remains as a memory of rotation. The stress in the circumferential direction increases after the rotation, whereas that in the radial direction decreases.
	    The residual stress is analytically related to the plastic deformation induced by the rotation.
	    Based on the time evolution of plastic deformation, we theoretically describe the mechanism of the changes in the stress distribution.
		
		%The residual tension in paste after rotational vibration is investigated based on a three-dimensional elasto-plastic model. 
		%We numerically find that the residual tension increases along the rotational direction after the vibration due to the plastic deformation, which is consistent with crack patterns in the experiments of drying pastes.
		%We analytically derive linearized equations, which explain the increase of the residual tension due to the shear deformation. 
		% \begin{description}
			% Secondary publications and information retrieval purposes.
			% You may use the \texttt{description} environment to structure your abstract;
			% use the optional argument of the \verb+\item+ command to give the category of each item. 
			% \end{description}
	\end{abstract}
	
	\maketitle
	
	%\tableofcontents

	\section{\label{sec:level1}Introduction}
	
\begin{figure}[htbp]
		\begin{minipage}[b]{0.45\linewidth}
			\centering
			\includegraphics[keepaspectratio, height=3.5cm]{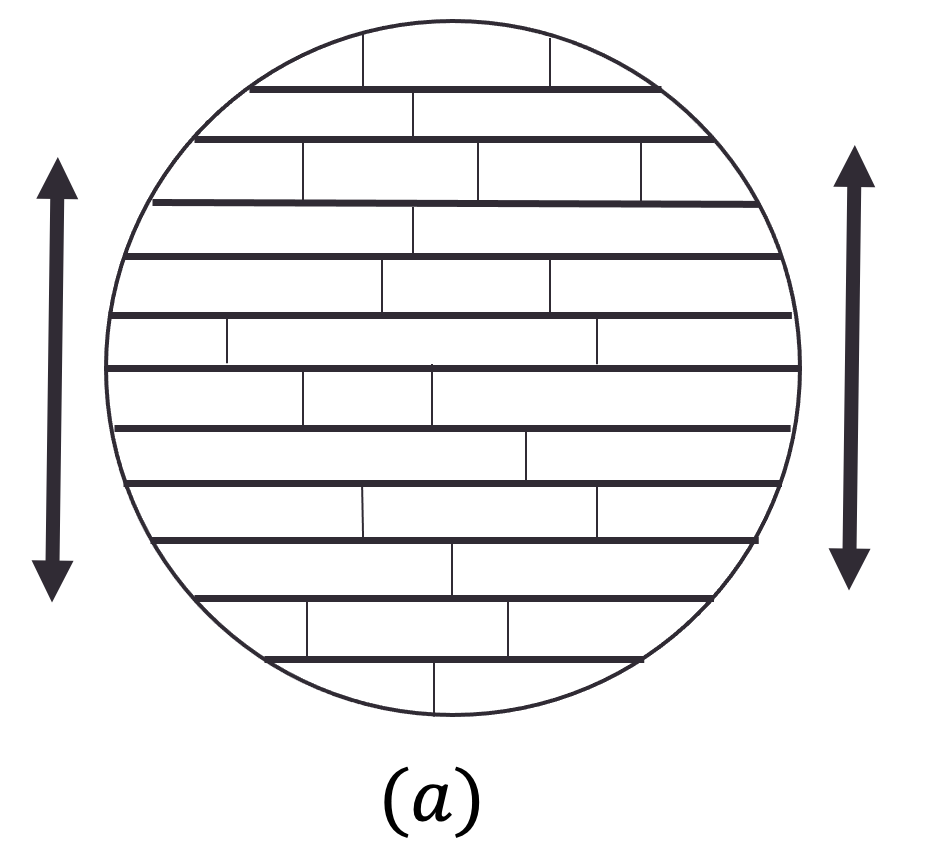}
			%    \subcaption{}
		\end{minipage}
		\begin{minipage}[b]{0.45\linewidth}
			\centering
			\includegraphics[keepaspectratio, height=3.5cm]{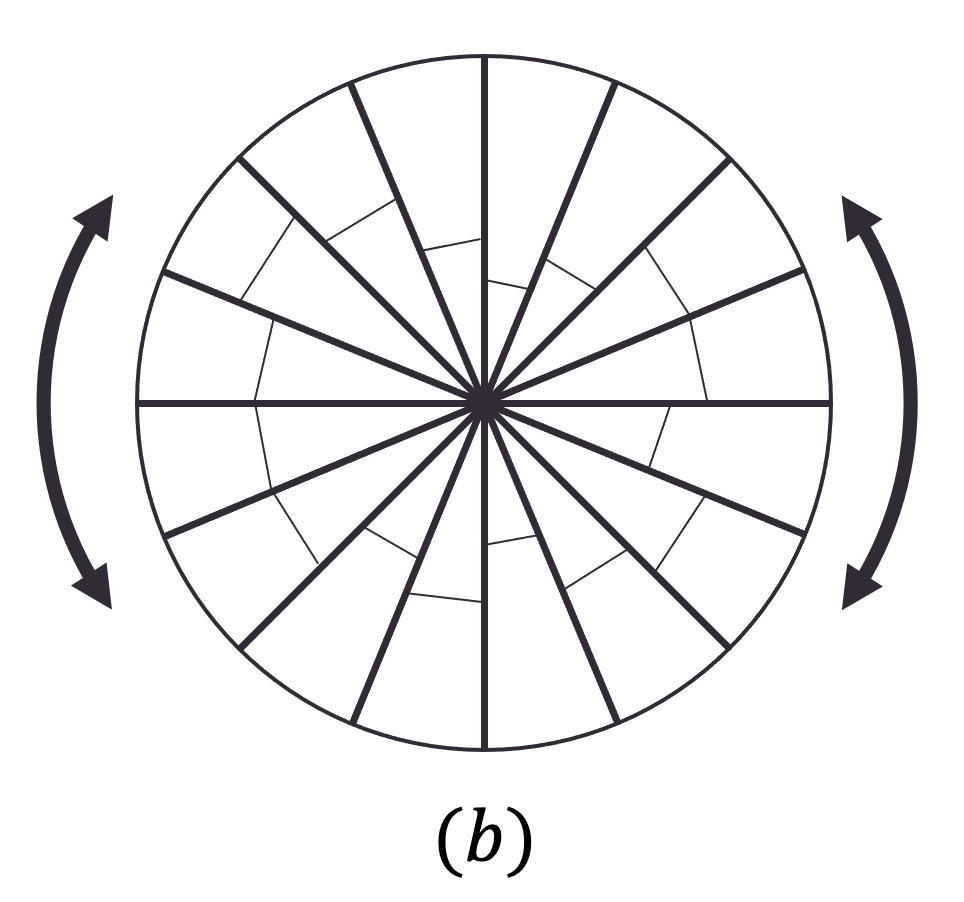}
			%    \subcaption{}
		\end{minipage}
		\caption{Schematics of crack patterns on the surface of a paste in a cylindrical container. (a) The lamellar crack pattern after oscillation in one direction. (b) 
			Radial crack pattern after oscillation in an angular direction.
			The thick lines represent the primary cracks formed initially.
			The thin lines represent the secondary cracks that are formed as desiccation progressed.
		}
		\label{fig1}
	\end{figure}

	When a paste comprising water and powder is dried in a shallow container, cracks appear on its surface \cite{Goehring2015}.
	Crack patterns induced by desiccation are typically random and isotropic \cite{Kindle1917,Groisman1994,Bohn2005,Shorlin2000,Goehring2010}. 
	However, recent experiments have revealed that patterns become anisotropic when external fields are applied upon them \cite{Nakahara2005,Nakahara2006a,Nakahara2006b,Matsuo2012,Nakayama2013,Mal2005,Khatun2013,Pauchard2008,Ngo2008,Lama2016,Szatmari2021,Uemura2024,Baba2023,Nakahara2011}.
	In Refs. \cite{Nakahara2005,Nakahara2006a,Nakahara2006b}, it has been  reported that the crack patterns on a paste made of calcium carbonate (\ce{CaCO3}) are controlled by horizontally oscillating the container before desiccation. 
	When the paste is oscillated in a cylindrical container in one direction before desiccation, cracks perpendicular to the direction of the oscillation are formed, resulting in the lamellar crack pattern shown in Fig. \ref{fig1}~(a) \cite{Nakahara2005}. 
	In contrast, when the container was rotated horizontally in the angular direction, a radial crack pattern, as shown in Fig. \ref{fig1}~(b), is formed.
	Cracks are formed days after the oscillation, indicating that the paste memorizes the direction of the oscillation.
	
	Pastes are elasto-plastic material characterized by the yield stress.
	According to Ref. \cite{Nakahara2005}, the memory effect occurs only when the stress induced by the oscillation exceeds the yield stress of the paste and causes plastic deformation.
	Therefore, the experimental results in Ref. \cite{Nakahara2005} suggest that the memory effect is related to the plastic deformation induced by oscillation.
	In fact, plastic deformation has been visualized in the paste showing the memory effect \cite{Nakahara2019}.

	Cracks are formed when the tension in the paste is increased by desiccation and the fracture criterion is satisfied \cite{Goehring2015,Kitsunezaki1999,Shingh2007,Man2008,Goehring2010,Halasz2017,Ito2014a,Ito2014b,Hirobe2016,Hirobe2017a,Hirobe2017b,Hirobe2019}.
	Hence, the memory effect is considered to result from a change in the tension or criterion.
	Recent experiments have revealed that desiccation does not change the criterion \cite{Kitsunezaki2016, Kitsunezaki2017}. 
	This indicates that the tension induced by plastic deformation leads to the memory effect on the crack patterns.
	Therefore, understanding the mechanism of the residual stress after oscillation clarifies the origin of the memory effect.

	In this respect, elasto-plastic models have been proposed to relate the residual stress to plastic deformation due to external oscillation in Refs. \cite{Otsuki2005,Ooshida2008,Ooshida2009,Morita2021}.
	These models are used to study the plastic deformation of pastes after oscillation in one direction. 
	In these models, the residual tension in the direction of oscillation increases owing to plastic deformation.
	Residual tension is enhanced by desiccation, creating lamellar crack patterns, as shown in Fig. \ref{fig1}(a) \cite{Otsuki2005}.
	However, the models in the previous studies have been applied only to two-dimensional deformation in the vertical plane corresponding to lamellar crack patterns.
	Therefore, the mechanism for residual stress after horizontal rotation that leads to the radial crack pattern, as shown in Fig. \ref{fig1}(b), remains unclear.

	In this study, we investigate the stress distribution of a paste in a cylindrical container rotated horizontally.
	We explain our setup and model in Sec. \ref{Sec:Model}. 
	In Sec. \ref{Sec:Evolve}, we introduce an elasto-plastic model for the paste.
	The boundary conditions and the choice of parameters are presented in Sec. \ref{Sec:BC}.
	In Sec. \ref{Sec:Result}, we discuss residual stress after rotation.
	In Sec. \ref{Sec:Simulation}, we show our numerical results, where the residual tension increases in the circumferential direction.
	Based on the time evolution of plastic deformation, we theoretically reveal the mechanism of increasing residual tension in Sec. \ref{Sec:Analysis}.
	Finally, we conclude the study and discuss the results in Sec. \ref{Sec:Summary}.
	%In Sec. \ref{App:B}, we derive the equation relating the left Cauchy-Green tensor and the plastic deformation.
	%In Sec. \ref{App:DG}, we derive the time evolution equations for the plastic deformation in cylindrical coordinates.

	\section{Setup}
	\label{Sec:Model}

	%{\bf NOTE: We use $dz=0.4$, but $dz=0.8$ gives quantitatively identical results.}
	
	In this section, we first present the time evolution equations for a paste in a cylindrical container. We then present the boundary conditions and parameters for the numerical simulation.

	\subsection{Elasto-plastic model}
	\label{Sec:Evolve}
	
\begin{figure}[h]
		\includegraphics[width=8cm]{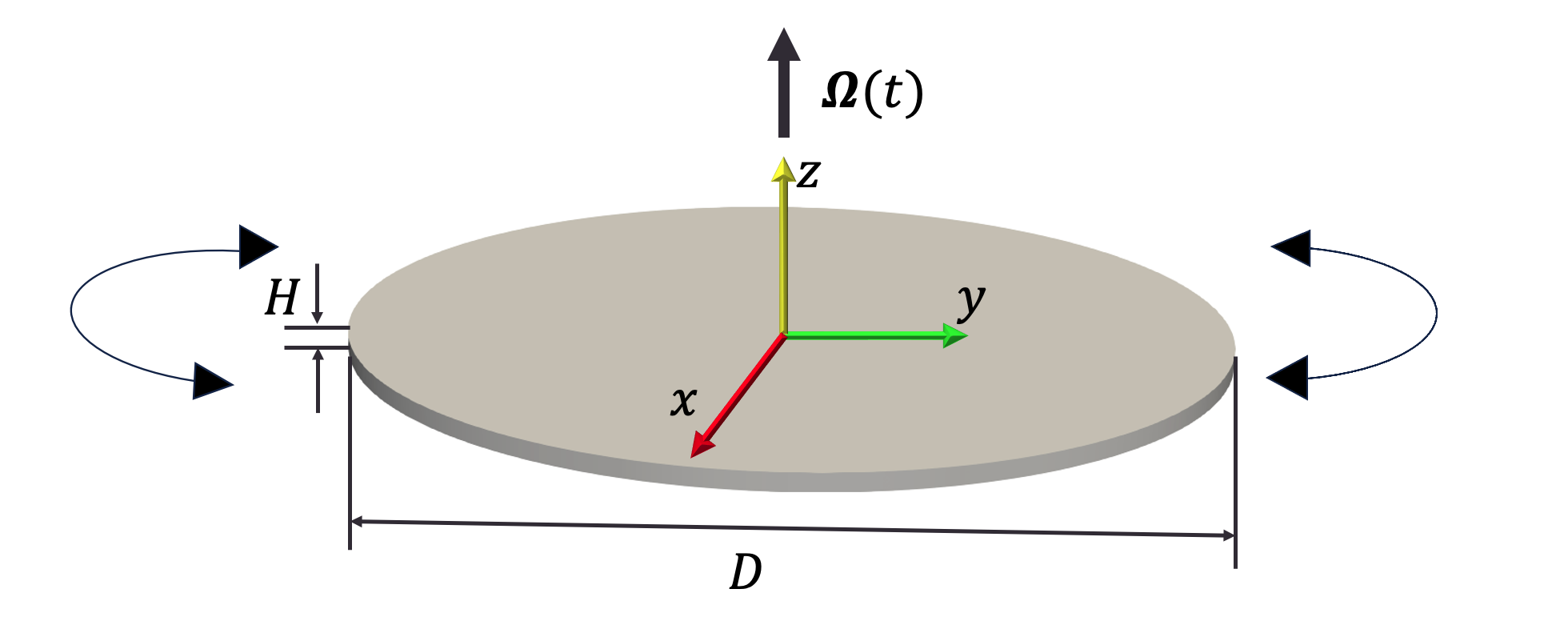}
		\caption{Schematic of a paste in a cylindrical container rotated with angular velocity $\bm{\Omega}(t)$.}
		\label{Fig:setup} 
	\end{figure}

	We consider a 3D elasto-plastic paste with a thickness of $H$ in a cylindrical container with a diameter of $D$ as shown in Fig. \ref{Fig:setup}.
	Cartesian coordinates $(x,y,z)$ are fixed to the container.
	The centre of the container is at $(x,y)=(0, 0)$ and the bottom at $z=0$.
	The container is rotated at an angular velocity of $\bm{\Omega}(t)$.
	
	The configuration of the paste is represented by the current coordinates $\bm{r}$ mapped from the initial coordinates $\bm{X}=(X, Y, Z)$ as follows: 
	\begin{align}
		\label{Eq:coordinate}
		\bm{r}(\bm{X},t)=\left[
		\begin{array}{c}
			x(\bm{X},t)\\
			y(\bm{X},t)\\
			z(\bm{X},t)
		\end{array}
		\right]_{\mathrm{C}}.
	\end{align}
	where $[\cdot ]_{\mathrm{C}}$ denotes the representation in Cartesian coordinates.
	Configuration $\bm{r}(\bm{X},t)$ satisfies
	\begin{align}
		\label{Eq:initial}
		\bm{r}(\bm{X},t=0)=\bm{X}.
	\end{align}

	The time evolution equation is given by \cite{Marsden1994,Romano2014,Truesdell2004,Beatty1987}
	\begin{eqnarray}
		\label{Eq:time}
		\rho\frac{\partial^{2}\bm{r}}{\partial t^{2}}&=&\mathrm{div}\bm{\sigma}+\bm{f},
	\end{eqnarray}
	where $\rho$ denotes the density of the paste, $\bm{f}$ is the body force, and $\bm{\sigma}$ is the Cauchy stress tensor.
	Here, $\bm{\sigma}$ consists of an elasto-plastic part $\bm{\sigma}^{\mathrm{(EP)}}$ and a viscous part $\bm{\sigma}^{(\mathrm{V})}$ as follows:
	\begin{align}
		\bm{\sigma}=
		\bm{\sigma}^{(\mathrm{EP})}+\bm{\sigma}^{(\mathrm{V})}.
		\label{Eq:sigma}
	\end{align}
	The body force $\bm{f}$ is generated by the rotation of the container as follows:
	\begin{align}
		\label{Eq:force}
		\bm{f}=-\rho\bm{\Omega}\times(\bm{\Omega}\times\bm{r})-2\rho\bm{\Omega}\times\dot{\bm{r}}-\rho\dot{\bm{\Omega}}\times\bm{r},
	\end{align}
	where the first, second, and third terms represent the centrifugal, Coriolis, and Euler forces, respectively.

	According to Ref. \cite{Morita2021}, we assume that the elasto-plastic part $\bm{\sigma}^{\mathrm{(EP)}}$ is given by
	\begin{eqnarray}
		\label{Eq:sigmaEP}
		\bm{\sigma}^{(\mathrm{EP})}&=&\{W+I_{3}H'(I_{3})+\mu\}\openone+\mu(\bm{B}-\openone)
	\end{eqnarray}
	with the left Cauchy-Green strain tensor $\bm{B}$.
	Here, $W$ is the Hadamard strain energy \cite{Destrade2005}
	\begin{eqnarray}
		\label{Eq:W}
		W&=&\frac{1}{2} \left \{ \mu(I_{1}-3) + H \right \}
	\end{eqnarray}
	with
	\begin{eqnarray}
		\label{Eq:H}
		H&=&(\lambda+\mu)(I_{3}-1)+2(\lambda+2\mu)(\sqrt{I_{3}-1}),
	\end{eqnarray}
	where $\lambda$ and $\mu$ are the Lam\'{e}s constants.
	Note that $I_{1}=\mathrm{tr}\bm{B}$ and $I_{3}= \mathrm{det}\bm{B}$ are the rotation invariants of the left Cauchy-Green strain tensor.
	We assume $\lambda/\mu \gg 1$ to effectively realize the incompressibility of the paste.
	The viscous part, $\bm{\sigma}^{(\mathrm{V})}$, is given by
	\begin{eqnarray}
		\label{Eq:sigmaV}
		\bm{\sigma}^{(\mathrm{V})}=2\eta\bm{S}
	\end{eqnarray}
	with the viscosity $\eta$ and the stretching tensor $\bm{S}$ \cite{Romano2014}.

	The left Cauchy-Green tensor is defined as \cite{Romano2014,Truesdell2004}
	\begin{align}
		\label{Eq:B_F}
		\bm{B}=\bm{F}\bm{F}^{\rm T}
	\end{align}
	with the deformation gradient tensor $\bm{F}$. 
	The deformation gradient tensor $\bm{F}$ represents the transformation of the infinitesimal line element $d \bm{r}$ between the two points labeled as $\bm{X}$ and $\bm{X} + d \bm{X}$ from its local stress-free ``natural state'' $d \bm{r}^\natural$ as \cite{Morita2021}
	\begin{align}
		\label{Eq:F_dr}
		d\bm{r}=\bm{F} d \bm{r}^\natural.
	\end{align}
	The square of the distance between the two points is given by
	\begin{align}
		\label{Eq:gij}
		|d\bm{r}|^2  =g_{ij}dX^{i}dX^{j}, \quad
		g_{ij}=(\partial_{i}\bm{r})\cdot(\partial_{j}\bm{r}),
	\end{align}
	where the metric tensor is denoted by $(g_{ij})$ or $\bm{g}$.
	Here, we use $\partial_i = \partial / \partial X^i$ and the Einstein notation for the summation of the repeated indices.
	The square of the ``natural distance'' between the two points is given by
	\begin{align}
		|d\bm{r}^\natural|^2 = g_{ij}^\natural d X^i dX^j,
	\end{align}
	where $(g_{ij}^\natural)$ or $\bm{g}^\natural$ is the natural metric tensor satisfying $g_{ij}^\natural = g_{ji}^\natural$.
	
	The left Cauchy-Green tensor is represented by \cite{Morita2021}
	\begin{align}
		\label{Eq:B}
		\bm{B}=g_{\natural}^{ij}(\partial_{i}\bm{r})\otimes(\partial_{j}\bm{r}),
	\end{align}
	where $(g_\natural^{ij})$ or $\bm{g}_\natural$ denotes the inverse of $\bm{g}^\natural$ satisfying
	\begin{align}
		\label{Eq:gij_nat_inv}
		\bm{g}^\natural \bm{g}_\natural = \openone.
	\end{align}
	The stretching tensor $\bm{S}$ is given by \cite{Morita2021}
	\begin{align}
		\label{Eq:D}
		\bm{S}=-\frac{1}{2}(\partial_t g^{ij}) \left\{(\partial_{i}\bm{r})\otimes(\partial_{j}\bm{r})\right\},
	\end{align}
	where $(g^{ij})$ or $\bm{g}^{-1}$ is the inverse of $\bm{g}$ satisfying
	\begin{align}
		\label{Eq:gij_inv}
		\bm{g} \bm{g}^{-1} = \openone.
	\end{align}
	
	Plastic deformation is characterized by the natural metric tensor. 
	The natural metric tensor $g_{ij}^{\natural}$ is initially set to $g_{ij}^{\natural} =  \delta_{ij}$ at $t=0$ and evolves when the equivalent stress $\bar{\sigma}$ exceeds the yield stress $\sigma_{\rm Y}$. 
	Following Refs. \cite{Ooshida2008,Ooshida2009}, the time evolution of $g^{ij}_{\natural}$ is given by
	\begin{eqnarray}
		\label{Eq:evolve_g}
		\tau \partial_{t} g_{\natural}^{ij} =  {\Gamma}g^{ij}-g_{\natural}^{ij} 
	\end{eqnarray}
	with a relaxation time $\tau$ and constant $\Gamma$, which is given by
	\begin{eqnarray}
		\label{Eq:Gamma}
		\Gamma=\frac{3}{g^{ij} g^{\natural}_{ij}}
	\end{eqnarray}
	to satisfy the incompressibility condition ${\rm det} \bm {g}^\natural=1$.
	The relaxation time is given by
	\begin{eqnarray}
		\label{Eq:tau}
		\tau^{-1}\left ( \bar{\sigma} \right) = \frac{\mu}{\eta_{\rm p}} \mathrm{max} \left ( 0, 1 - \frac{\sigma_{\rm Y}}{ \bar{\sigma}} \right )
	\end{eqnarray}
	with a constant $\eta_{\rm p}$ \cite{Jones2009}.
	The equivalent stress is given by
	\begin{align}
		\label{Eq:bar_sigma}
		\bar{\sigma}^{2}&=\frac{1}{2}\left(\sigma_{xx}-\sigma_{yy}\right)^{2} \nonumber \\
		&+\frac{1}{2}\left(\sigma_{yy}-\sigma_{zz}\right)^{2} \nonumber \\
		&+\frac{1}{2}\left(\sigma_{zz}-\sigma_{xx}\right)^{2} \nonumber \\ &+3\left(\sigma_{xy}^{2}+\sigma_{yz}^{2}+\sigma_{zx}^{2}\right),
	\end{align}
	where $\sigma_{i j}$ with $i, j = x, y, z$ is a component of the stress tensor.
	Equation \eqref{Eq:evolve_g} with Eq. \eqref{Eq:tau} indicates that the plastic deformation associated with the change in $(g_{ij}^\natural)$ occurs due to the finite relaxation time $\tau$ when the von Mises yield criterion $\bar{\sigma} \ge \sigma_{\rm Y}$ is satisfied.

	\subsection{Boundary conditions and parameters}
	\label{Sec:BC}
	
	We introduce the cylindrical coordinates, where $\bm{X}$ and $\bm{r}$ are expressed as follows:
	\begin{align}
		\label{Eq:X_cy}
		\bm{X}=\left[
		\begin{array}{c}
			R\cos\Theta\\
			R\sin\Theta\\
			Z
		\end{array}
		\right]_{\mathrm{C}}, \quad
		\bm{r}=\left[
		\begin{array}{c}
			r\mathrm{cos}\theta\\
			r\mathrm{sin}\theta\\
			z
		\end{array}
		\right]_{\mathrm{C}}
	\end{align}
	with 
	\begin{align}
		\label{Eq:R_Theta}
		R = \sqrt{X^2+Y^2}, \quad \Theta = \tan^{-1} \left ( \frac{Y}{X} \right )
	\end{align}
	and
	\begin{align}
		\label{Eq:r_theta}
		r = \sqrt{x^2+y^2}, \quad \theta = \tan^{-1} \left ( \frac{y}{x} \right ).
	\end{align}
	We also introduce unit vectors 
	\begin{align}
		\bm{e}_r=\left[
		\begin{array}{c}
			\cos \theta \\
			\sin \theta \\
			0
		\end{array}
		\right]_{\mathrm{C}},
		\ 
		\bm{e}_\theta=\left[
		\begin{array}{c}
			-\sin \theta \\
			\cos \theta \\
			0
		\end{array}
		\right]_{\mathrm{C}},
		\
		\bm{e}_z=\left[
		\begin{array}{c}
			0 \\
			0 \\
			1
		\end{array}
		\right]_{\mathrm{C}}.
	\end{align}

	We assume the no-slip boundary conditions $\partial_t \bm{r}(\bm{X}, t)= \bm{0}$ at the bottom of the container ($Z=0$).
	The stress applied to the free surface at $Z=H$ is given by $\bm{\sigma} \bm{n} = \bm{0}$, where $\bm{n}$ denotes the normal unit vector of the surface.
	At the lateral wall ($R=D/2$), we assume $\partial_t \bm{r}(\bm{X}, t) \cdot \bm{e}_r = 0$,
	$\bm{e}_\theta \cdot \bm{\sigma} \bm{e}_r =  0$, and $\bm{e}_z \cdot \bm{\sigma} \bm{e}_r  =0$, where the paste does not leave the wall and slides freely along the wall.
	
	We oscillate the container $N$ times in the angular direction with period $T$ and relax the paste during a relaxation time $T_{\rm R}$ from the initial state.
	The angular velocity $\bm{\Omega}(t)$ is given by
	\begin{align}
		\label{Eq:Omega}
		\bm{\Omega}=\left[
		\begin{array}{c}
			0\\
			0\\
			\frac{d}{dt}\Phi(t)
		\end{array}
		\right]_{\rm C}
	\end{align}
	with the rotation angle 
	\begin{align}
		\label{Eq:Phi}
		\Phi(t) = 
		\begin{cases}
			A\mathrm{cos}\left(\frac{2\pi t}{T}\right) - 1& (0\leq t < NT)\\
			0 & (NT \leq t \leq NT+T_{\rm R})
		\end{cases}
	\end{align}
	and the amplitude $A$.
	We study the stress distribution after the rotation at $t= NT + T_{\rm R}$.

	We adopt $D/H = 50.0$, $\lambda/\mu = 1.0 \times 10^{5}$, $\eta/(\mu/\tau_0) =  1.0$, $\eta_{\rm P} =  2.0 \tau_0^{-1}$, and $\sigma_{Y}/\mu = 0.8$ with the unit of time $\tau_0=H\sqrt{\rho/\mu}$ following the previous experiments \cite{Nakahara2005} and simulations \cite{Morita2021}.
	We set $T/\tau_0= 5.0$, $ A = 0.1$, $N = 10$, and $T_{\rm R}/\tau_0 = 80.0$.
	Equation \eqref{Eq:time} is solved numerically by using the finite element method with a time step $\Delta t/\tau_0 = 2.0 \times 10^{-4}$.
	The paste is divided into cubes of length $\Delta x/D = 1.0 \times 10^{-2}$ consisting of six tetrahedrons.
	Note that we have checked that $N\ge10$ and a longer $T_{\rm R}$ give the same results shown below.
	%We use $\Delta t/\tau_0 = 2.0 \times 10^{-4}$ and $\Delta x/L = 5\times10^{-3}$, which corresponds to the number of nodes $N_{\mathrm{nodes}} =  159085$.

	\section{Residual stress after the rotation}
	\label{Sec:Result}
	
	In this section, we first present the numerical results of the stress distribution after the rotation.
	Then, the mechanism of the residual stress is theoretically explained.

	\subsection{Distribution of the residual stress}
	\label{Sec:Simulation}
	
\begin{figure}[htbp]
		\centering
		\includegraphics[keepaspectratio, width=8cm]{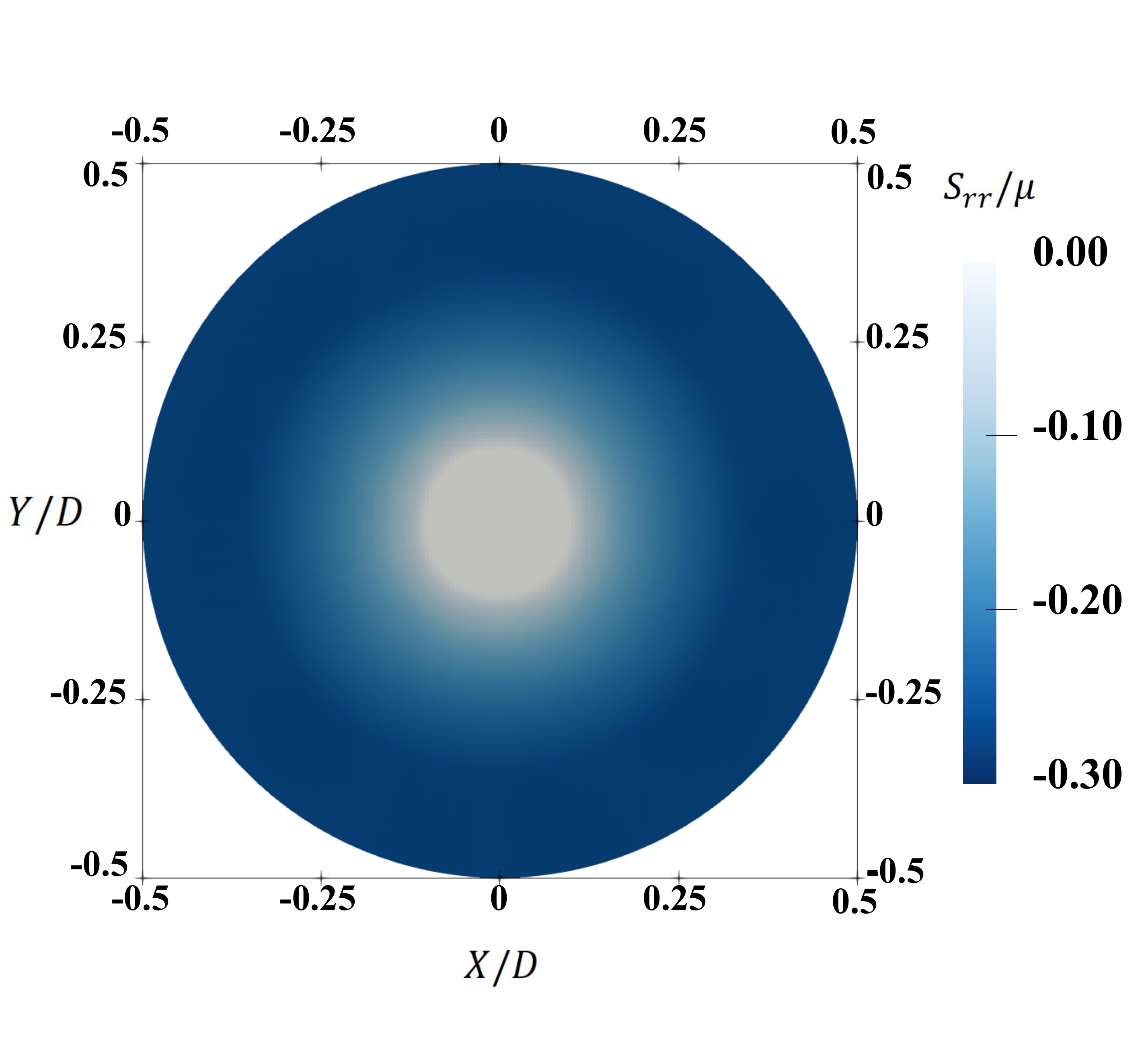}
		\caption{Spatial distribution of $S_{rr}$ at $Z=0$ after the rotation.}
		\label{fig:Sr} 
	\end{figure}

\begin{figure}[htbp]
		\centering
		\includegraphics[keepaspectratio, width=8cm]{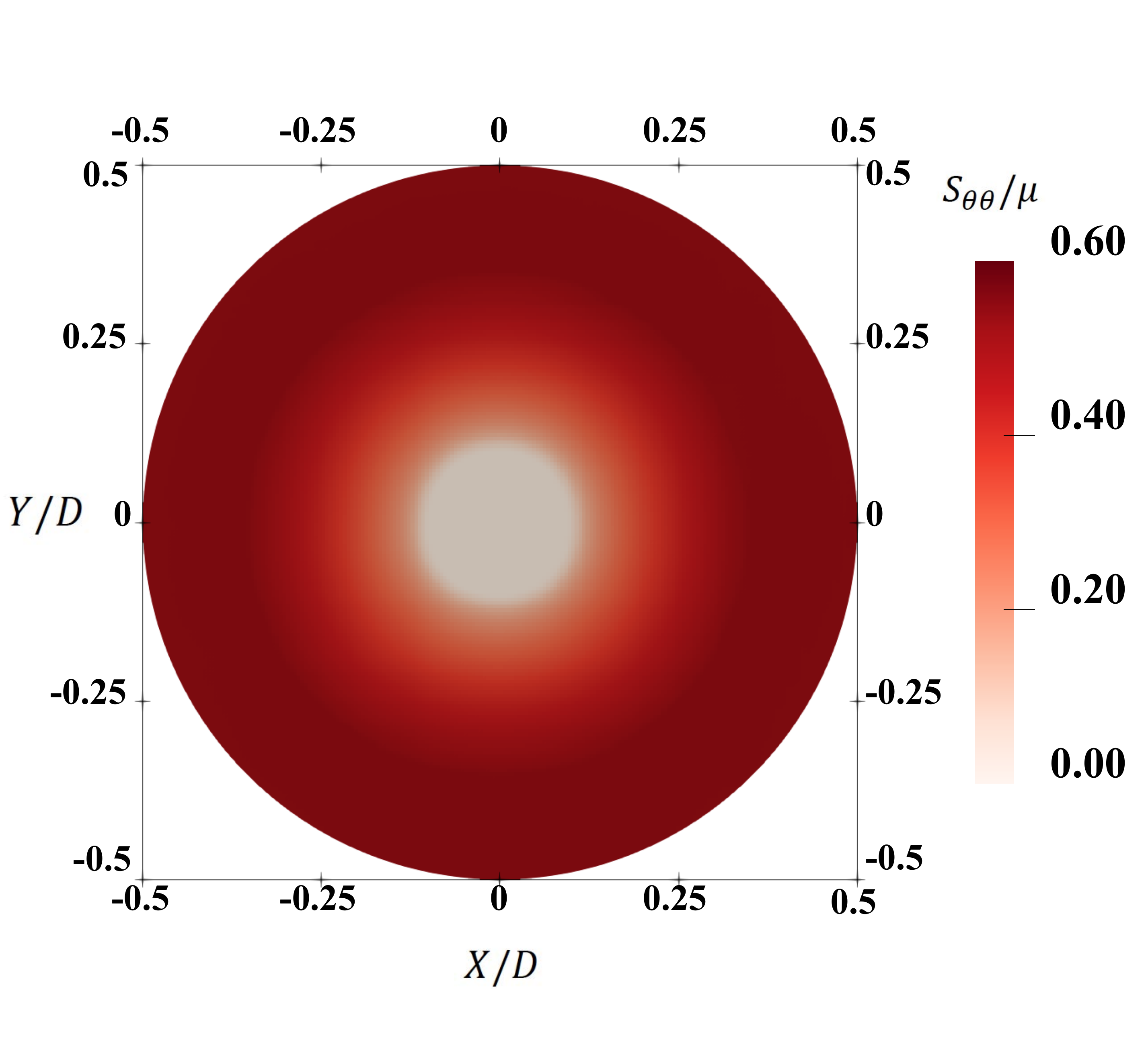}
		\caption{Spatial distribution of $S_{\theta \theta}$ at $Z=0$ after the rotation. }
		\label{fig:St} 
	\end{figure}

	We focus on the deviatoric stress
	\begin{align}
		\label{Eq:S}
		\bm{S}=\bm{\sigma}-\left(\frac{\mathrm{tr}\bm{\sigma}}{3}\right)\bm{I}
	\end{align}
	because a crack is formed when $\bm{S}$ becomes significant.
	We introduce the components of the deviatoric stress tensor in the cylindrical coordinates as follows:
	\begin{align}
		\label{Eq:S_cy_com}
		S_{\alpha \beta}  = \bm{e}_\alpha \cdot \bm{S} \bm{e}_\beta
	\end{align}
	with $\alpha, \beta = r, \theta, z$ and plot the spatial distributions of $S_{rr}$ and $S_{\theta \theta}$ after rotation at the bottom $(Z=0)$ in Figs. \ref{fig:Sr} and \ref{fig:St}, respectively.
	The stress distributions exhibit rotational symmetry.
	$S_{rr}$ and $S_{\theta \theta}$ are $0$ near the centre of the system.
	However, in the outer region, the nonzero residual stress ($S_{rr}<0$ and $S_{\theta \theta}>0$) remains as a memory of the rotation even though the system is sufficiently relaxed after rotation.
	The magnitudes of $S_{rr}$ and $S_{\theta \theta}$ increase with increasing $R$.
	The components $S_{rr}$ and $S_{\theta \theta}$ represent the tension in the radial and circumferential directions, respectively.
	A positive value of $S_{\theta \theta}$ is consistent with the crack patterns shown in Fig. \ref{fig1}(b), where the radial cracks appear to release the increasing tension in the circumferential direction.
	
	%The negative normal component $S_{rr}$ in the radial direction (Fig. \ref{fig:Sr} indicates that the paste is compressed in this direction, while $S_{\theta \theta}>0$ (Fig. \ref{fig:St} indicates the paste is stretched in the rotational direction.
	%The crack is formed to release the increased tension.
	%Therefore, positive $S_{\theta \theta}$ results in the crack in the radial direction to decrease $S_{\theta \theta}$ when it is enhanced by the desiccation, which is consistent with the experiments [CITATION].

	%These values on the bottom of paste ($Z=0$) are shown in Fig \ref{fig:3.1a} and \ref{fig:3.1b}. $S_{ii}>0$ means that tension increases in i-direction, and $S_{ii}<0$ means that compression works in i-direction. $S_{ii}=0$ indicates that no stress remained. From Fig \ref{fig:3.1a}, which shows the distribution of $S_{rr}$, we can see that no radial stress around center is remained, and radial stress in the other area is under compression uniformly in the direction of rotation. On the other hand, from Fig \ref{fig:3.1b}, which shows the distribution of $S_{\theta\theta}$, we can see that no rotational stress around center is remained, and rotational stress in the other area is under tension uniformly in the direction of rotation. Therfore, we can see that tension on the bottom develops in rotational direction. We can expect that cracks induced by drying propagate in radial direction to release the stress developed in rotational direction. As expected, this result is consistent with the experimental result that radial cracks is formed.
	
\begin{figure}[htbp]
		\centering
		\includegraphics[keepaspectratio, width=8cm]{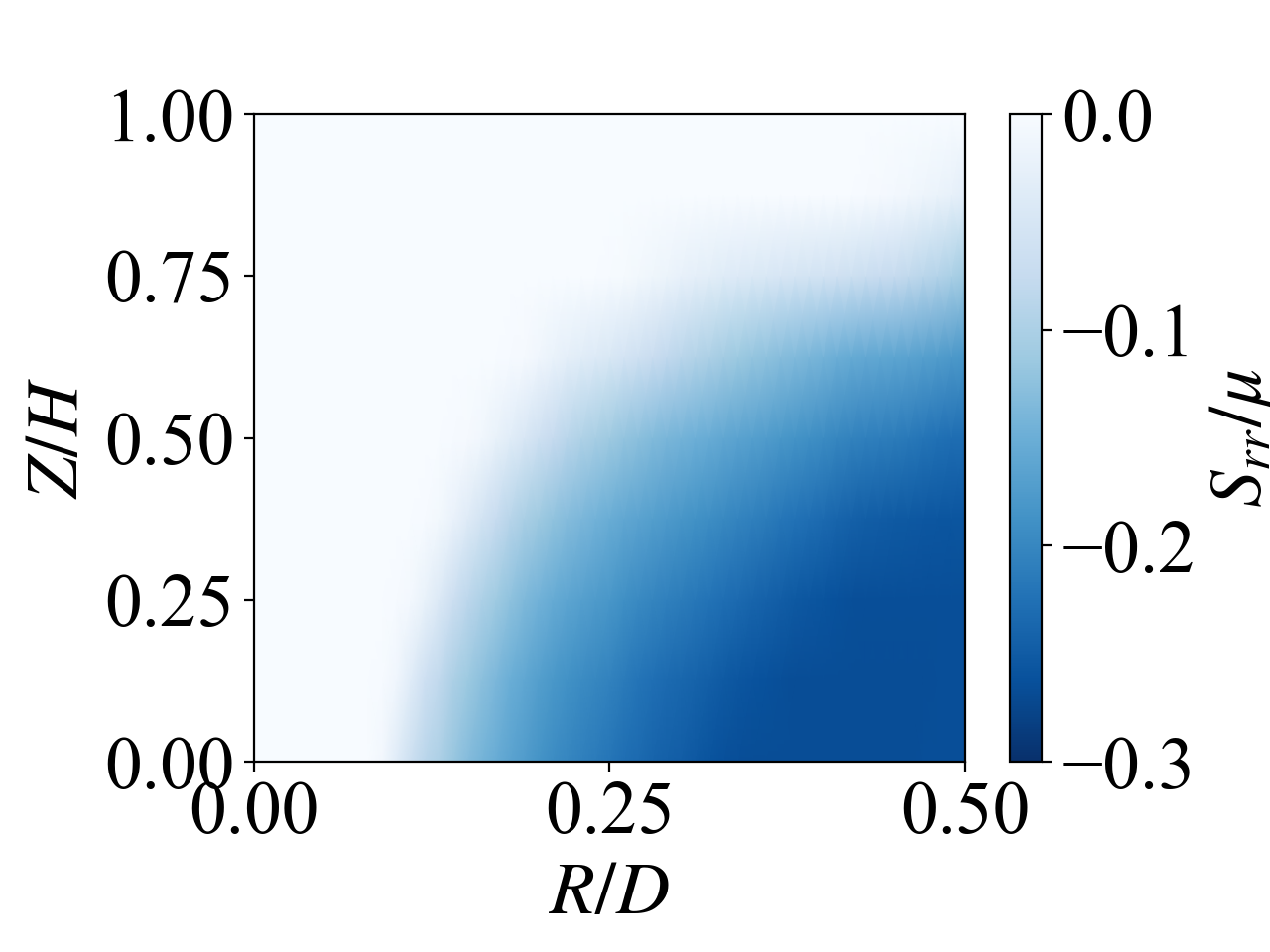}
		\caption{Spatial distribution of $S_{rr}$ for $\Theta = 0$ after the rotation.}
		\label{fig:Sr_RZ} 
	\end{figure}

\begin{figure}[htbp]
		\centering
		\includegraphics[keepaspectratio, width=8cm]{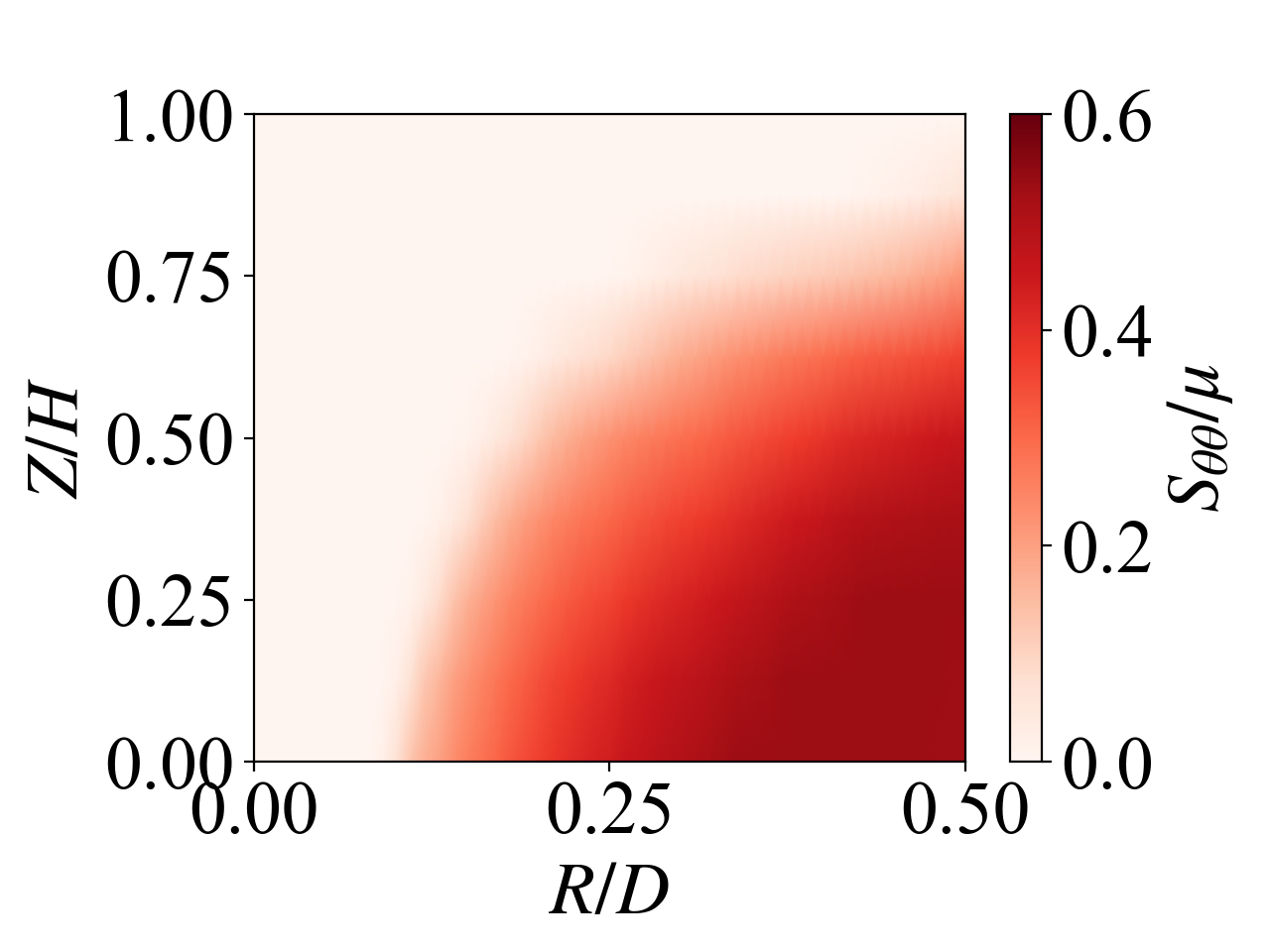}
		\caption{
			Spatial distribution of $S_{\theta \theta}$ for $\Theta = 0$ after the rotation.
		}
		\label{fig:St_RZ}
	\end{figure}

	In Figs. \ref{fig:Sr_RZ} and \ref{fig:St_RZ}, we plot the spatial distributions of $S_{rr}$ and $S_{\theta\theta}$ after the rotation for $\Theta=0$, respectively.
	At $R=D/2$ and $z=0$, $|S_{rr}|$ and $|S_{\theta\theta}|$ are largest.
	In this system, the Euler force along the circumferential direction is dominant in Eq. \eqref{Eq:force} and induces large deformation near the wall. 
	In addition, the shear stress becomes the largest near the bottom owing to the boundary conditions.
	Therefore, the equivalent stress becomes the largest at the bottom near the wall of the cylindrical container, where large plastic deformation and residual stress $S_{rr}<0$ and $S_{\theta\theta}>0$ occur.

	%We research how the residual stress in each direction changes by depth. Fig.\ref{fig:3.2} and Fig.\ref{fig:3.3} shows $S_{rr}$ and $S_{\theta\theta}$ by colormap, respectively, where the horizontal axis represents the distance in radial direction $R$, and the vertical axis represents the distance from the bottom of paste $Z$. From Fig.\ref{fig:3.2}, we can see that most part on the side of the wall near the bottom shows $S_{rr}<0$. This means that compression in radial direction develops near the bottom. Additionaly, from Fig.\ref{fig:3.2}, we can see that the higher the position is, the stronger the tention is. From Fig.\ref{fig:3.3}, we can see that most part on the side of the wall near the bottom shows $S_{\theta\theta}>0$. This means that tention in rotational direction develops near the bottom. Additionaly, from Fig.\ref{fig:3.3}, we can see that the higher the position is, the weaker the tention is.
	
	\subsection{Theoretical explanation for residual stress}
	\label{Sec:Analysis} 
	
	We theoretically analyze the origin of the negative $S_{rr}$ and positive $S_{\theta \theta}$ after the rotation.
	Substituting Eq. \eqref{Eq:sigma} into Eq. \eqref{Eq:S} with Eq. \eqref{Eq:sigmaEP}, we obtain
	\begin{align}
		\bm{S} = \mu \left ( \bm{B} - \frac{\mathrm{tr} \bm{B}}{3} \bm{I} \right ),
	\end{align}
	where we have neglected $\bm{\sigma}^{(\mathrm{V})}$ by assuming quasi-static deformation.
	By substituting this equation into Eq. \eqref{Eq:S_cy_com},
	$S_{rr}$ and $S_{\theta \theta}$ can be expressed as follows:
	\begin{align}
		S_{rr} & = \mu \left ( B_{rr} - \frac{B_{rr} + B_{\theta \theta} + B_{zz} }{3} \right ), \label{Eq:Sr_B}\\
		S_{\theta \theta } & = \mu \left ( B_{\theta \theta} - \frac{B_{rr} + B_{\theta \theta} + B_{zz} }{3} \right ) \label{Eq:St_B}
	\end{align}
	with the components of the left Cauchy-Green tensor in the cylindrical coordinates
	\begin{align}
		\label{Eq:B_cy}
		B_{\alpha \beta} = \bm{e}_\alpha \cdot  \bm{B} \bm{e}_\beta.
	\end{align}

	The left Cauchy-Green tensor is determined using the natural metric tensor $\bm{g}^\natural$, which is expressed as
	\begin{align}
		\label{Eq:g_G}
		\bm{g}^\natural = 
		\left(\begin{array}{ccc}
			g_{XX}^\natural&g_{XY}^\natural&g_{XZ}^\natural\\
			g_{XY}^\natural&g_{YY}^\natural&g_{YZ}^\natural\\
			g_{XZ}^\natural&g_{YZ}^\natural&g_{ZZ}^\natural
		\end{array}
		\right)
		=\bm{Q}^{\mathrm{T}} \bm{G}^\natural \bm{Q}
	\end{align}
	with the rotation matrix
	\begin{align}
		\label{Eq:Q}
		\bm{Q}=\left(
		\begin{array}{ccc}
			\cos\theta&\sin\theta&0\\
			-\sin\theta&\cos\theta&0\\
			0&0&1
		\end{array}\right).
	\end{align}
	Here, $\bm{G}^\natural = \bm{Q} \bm{g}^\natural \bm{Q}^{\mathrm{T}}$ is represented as
	\begin{align}
		\label{Eq:G_natural_app}
		\bm{G}^\natural
		=\left(
		\begin{array}{ccc}
			1+\Delta G_{rr}^\natural & \gamma & \delta\\
			\gamma & 1+\Delta G_{\theta \theta}^\natural & \epsilon \\
			\delta & \epsilon & \Xi
		\end{array}
		\right).
	\end{align}
	At $t=0$, $\Delta G_{rr}^\natural$, $\Delta G_{\theta \theta}^\natural $, $\gamma$, $\delta$, and $\epsilon$ are $0$, and plastic deformation is characterized by their time evolution.
	From the incompressible condition, $ \mathrm{det} \bm{g}^\natural= \mathrm{det} \bm{G}^\natural =1$, $\Xi$ satisfies
	\begin{align}
		\Xi  = \frac{1 + \epsilon^2 (1+\Delta G_{rr}^\natural )  + \delta^2 (1+\Delta G_{\theta \theta}^\natural ) - 2 \gamma \delta \epsilon}{ (1+\Delta G_{rr}^\natural )(1+\Delta G_{\theta \theta}^\natural )- \gamma^2}.
	\end{align}

	In our system, the Euler force causes the rigid rotation at each height $Z$.
	Therefore, we approximate the deformation of the paste as
	\begin{align}
		\label{Eq:deformation}
		r(\bm{X},t) = R, \ \theta(\bm{X},t) = \Theta + \varphi(Z,t), \ z(\bm{X},t) = Z
	\end{align}
	with a rotation angle $\varphi(Z,t)$.
	In the analysis below, we consider the case of a small deformation with $\varphi \ll 1$ and neglect the higher-order terms of $\Delta G_{rr}^\natural$, $\Delta G_{\theta \theta}^\natural $, $\gamma$, $\delta$, $\epsilon$, and $\bm{G}^\natural$.
	Using Eqs. \eqref{Eq:g_G}, \eqref{Eq:G_natural_app}, and \eqref{Eq:deformation} with Eq. \eqref{Eq:B}, we obtain
	\begin{align}
		\label{Eq:Br}
		& B_{rr} \simeq 1-\Delta G_{rr}^\natural, \\
		\label{Eq:Btheta}
		&  B_{\theta \theta} \simeq 1-\Delta G_{\theta \theta}^\natural,\\
		\label{Eq:Bz}
		& B_{zz} \simeq 1+\Delta G_{rr}^\natural+\Delta G_{\theta \theta}^\natural.
	\end{align}
	Appendix \ref{App:B} provides the derivation of these equations.
	Substituting Eqs. \eqref{Eq:Br}-\eqref{Eq:Bz} into Eqs. \eqref{Eq:Sr_B} and \eqref{Eq:St_B}, we obtain
	\begin{align}
		\label{Eq:S_rr_remain}
		S_{rr} \simeq -\mu \Delta G_{rr}^\natural, \\
		\label{Eq:S_tt_remain}
		S_{\theta \theta} \simeq -\mu \Delta G_{\theta \theta}^\natural.
	\end{align}
	These equations relate the plastic deformation characterized by $\Delta G_{rr}^\natural$ and $\Delta G_{\theta \theta}^\natural$ to the residual stress.

	Based on Eqs. \eqref{Eq:evolve_g}, \eqref{Eq:g_G}, \eqref{Eq:G_natural_app}, and \eqref{Eq:deformation}, we derive the time evolution equations of $\Delta G_{rr}^\natural$ and $\Delta G_{\theta \theta}^\natural$ as
	\begin{align}
		\label{Eq:evolve_Grr}
		\partial_{t}\Delta G_{rr}^\natural &\simeq \tau^{-1}\left \{ \frac{1}{3}R^2 (\partial_{Z}\varphi)^{2} -\Delta G_{rr}^\natural \right \}, \\
		\label{Eq:evolve_Gtt}
		\partial_{t}\Delta G_{\theta \theta}^\natural &\simeq \tau^{-1}\left \{ -\frac{2}{3} R^2 (\partial_{Z}\varphi)^{2} -\Delta G_{\theta \theta }^\natural \right \}.
	\end{align}
	See Appendix \ref{App:DG} for details on the derivation.
	As the shear strain represented by $R \left (\partial_{Z}\varphi \right )$ increases owing to the rotation, the equivalent stress $\bar{\sigma}$ exceeds $\sigma_{\rm Y}$ and $\tau^{-1}$ becomes nonzero, which causes a change in $\Delta G_{rr}^\natural$ and $\Delta G_{\theta \theta}^\natural$.
	In Eqs. \eqref{Eq:evolve_Grr} and \eqref{Eq:evolve_Gtt}, $\Delta G_{rr}^\natural$ converges to $\frac{1}{3}R^2 (\partial_{Z}\varphi)^{2}>0$, while $\Delta G_{\theta \theta}^\natural$ changes to $-\frac{2}{3}R^2 (\partial_{Z}\varphi)^{2}<0$.
	After the rotation of the cylinder stops, positive $\Delta G_{rr}^\natural$ and negative $\Delta G_{\theta \theta}^\natural$ remain because $\tau^{-1}$ in Eqs. \eqref{Eq:evolve_Grr} and \eqref{Eq:evolve_Gtt} become $0$ before the rotation stops, which leads to the residual stress $S_{rr}<0$ and $S_{\theta \theta}>0$ through Eqs. \eqref{Eq:S_rr_remain} and \eqref{Eq:S_tt_remain}.
	Note that Eq. \eqref{Eq:evolve_Gtt} corresponds to the time evolution equation of the natural metric tensor in pastes oscillated in one direction in Ref. \cite{Ooshida2008}.

\begin{figure}[htbp]
		\centering
		\includegraphics[keepaspectratio, width=8cm]{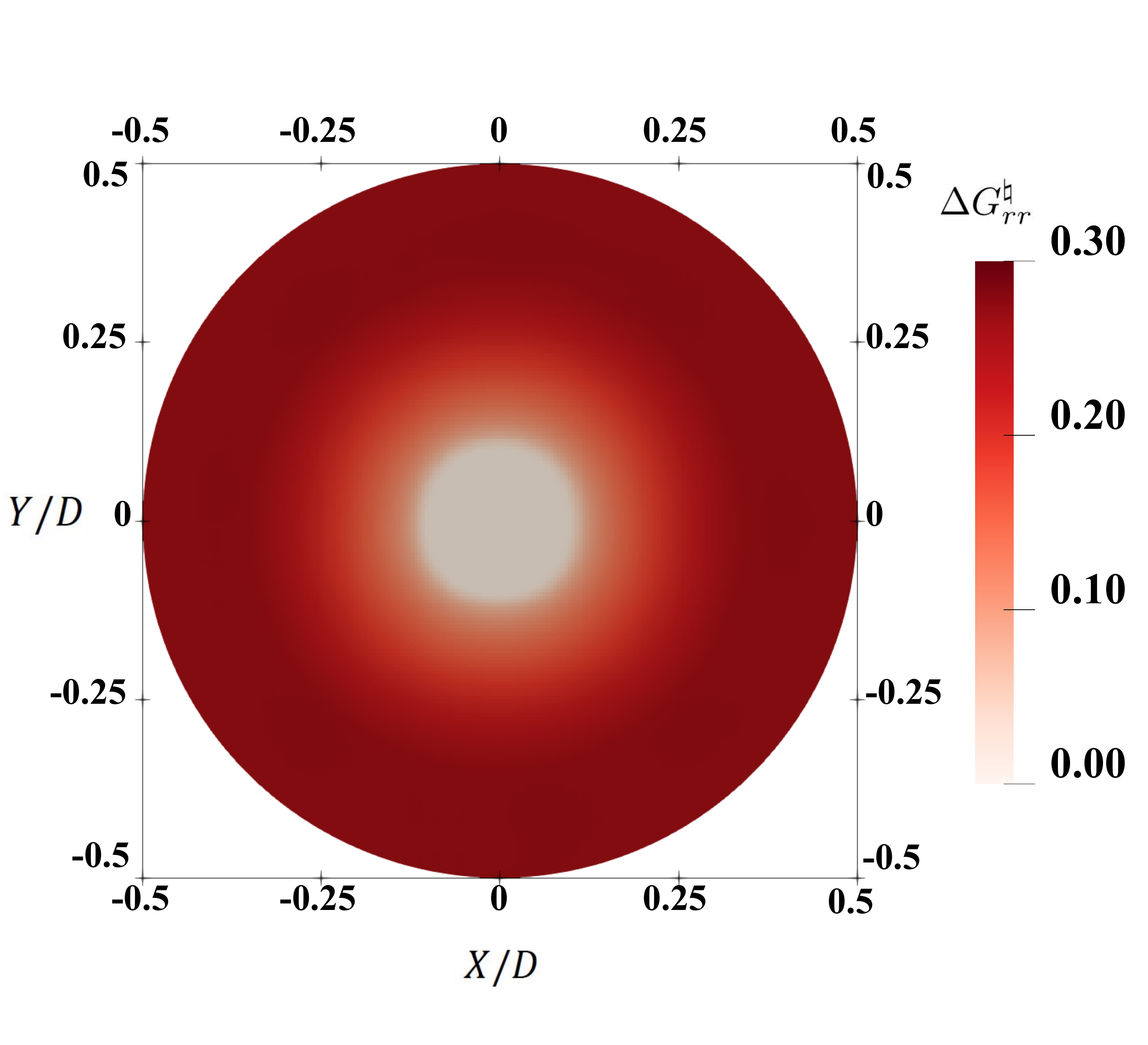}
		\caption{ 
			Spatial distribution of $\Delta G_{rr}^{\natural}$ at $Z=0$ after the rotation.
		}
		\label{fig:gnr}
	\end{figure}

\begin{figure}[htbp]
		\centering
		\includegraphics[keepaspectratio, width=8cm]{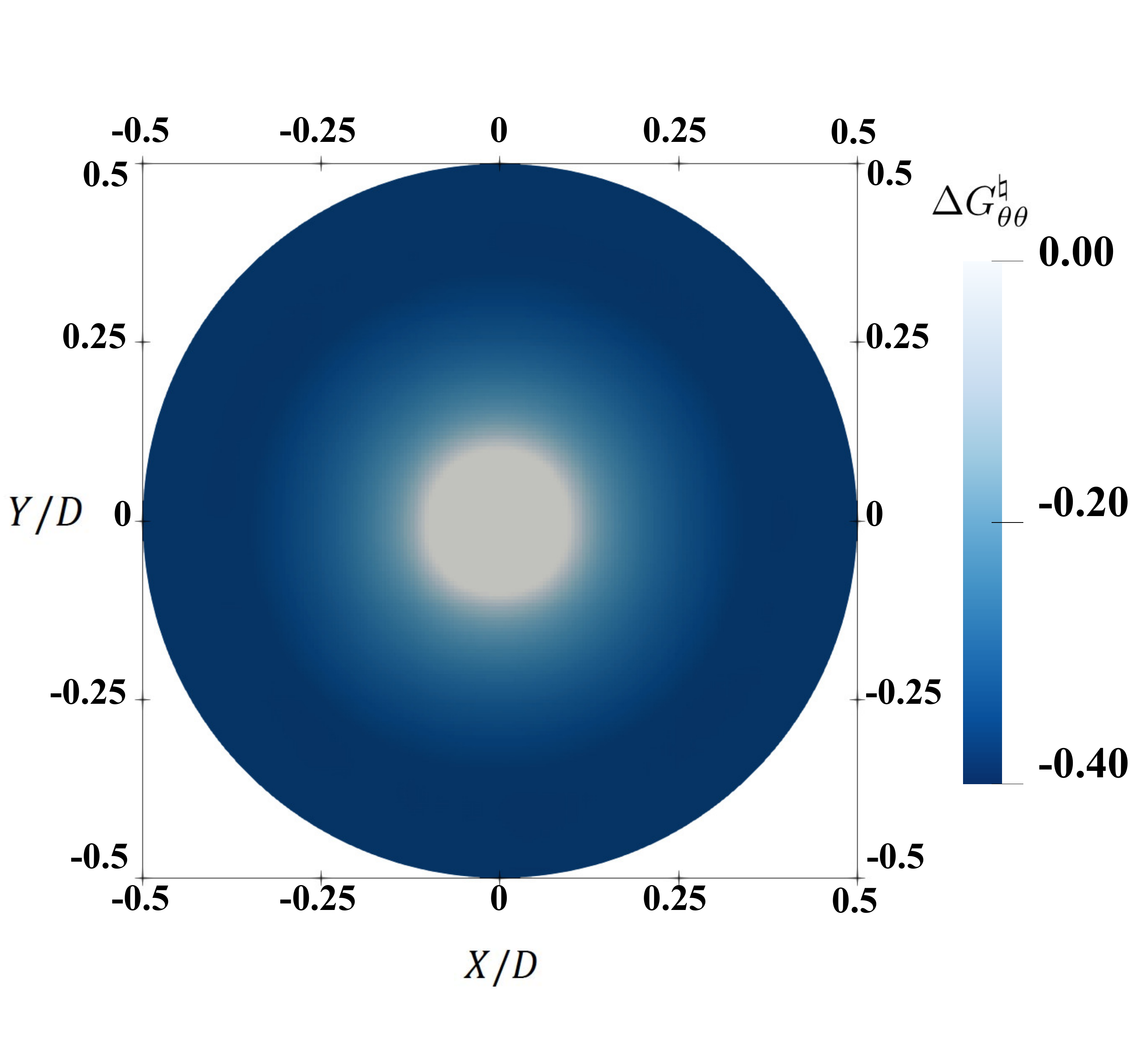}
		\caption{ 
			Spatial distribution of $\Delta G_{\theta \theta }^{\natural}$ at $Z=0$ after the rotation.
		}
		\label{fig:gnt}
	\end{figure}

	In Figs. \ref{fig:gnr} and \ref{fig:gnt}, we plot the spatial distributions of $\Delta G_{rr}^{\natural}$ and $\Delta G_{\theta \theta}^{\natural}$ at $Z=0$ after the rotation, respectively.
	The components $\Delta G_{rr}^{\natural}$ and $\Delta G_{\theta\theta}^{\natural}$ are $0$ near the center of the container.
	As $R$ increases, $\Delta G_{rr}^{\natural}$ increases from $0$ and $\Delta G_{\theta \theta}^{\natural}$ decreases.
	The numerical results for $\Delta G_{rr}^{\natural} \ge 0$ and  $\Delta G_{\theta \theta }^{\natural} \le 0$ are consistent with the time evolutions described by Eqs. \eqref{Eq:evolve_Grr} and \eqref{Eq:evolve_Gtt}.
	The spatial distributions of $\Delta G_{rr}^{\natural}$ and $\Delta G_{\theta \theta}^{\natural}$ are consistent with those of $S_{rr}$ and $S_{\theta \theta}$ in Figs. \ref{fig:Sr} and \ref{fig:St}, respectively.

\begin{figure}[htbp]
		\centering
		\includegraphics[keepaspectratio, width=8cm]{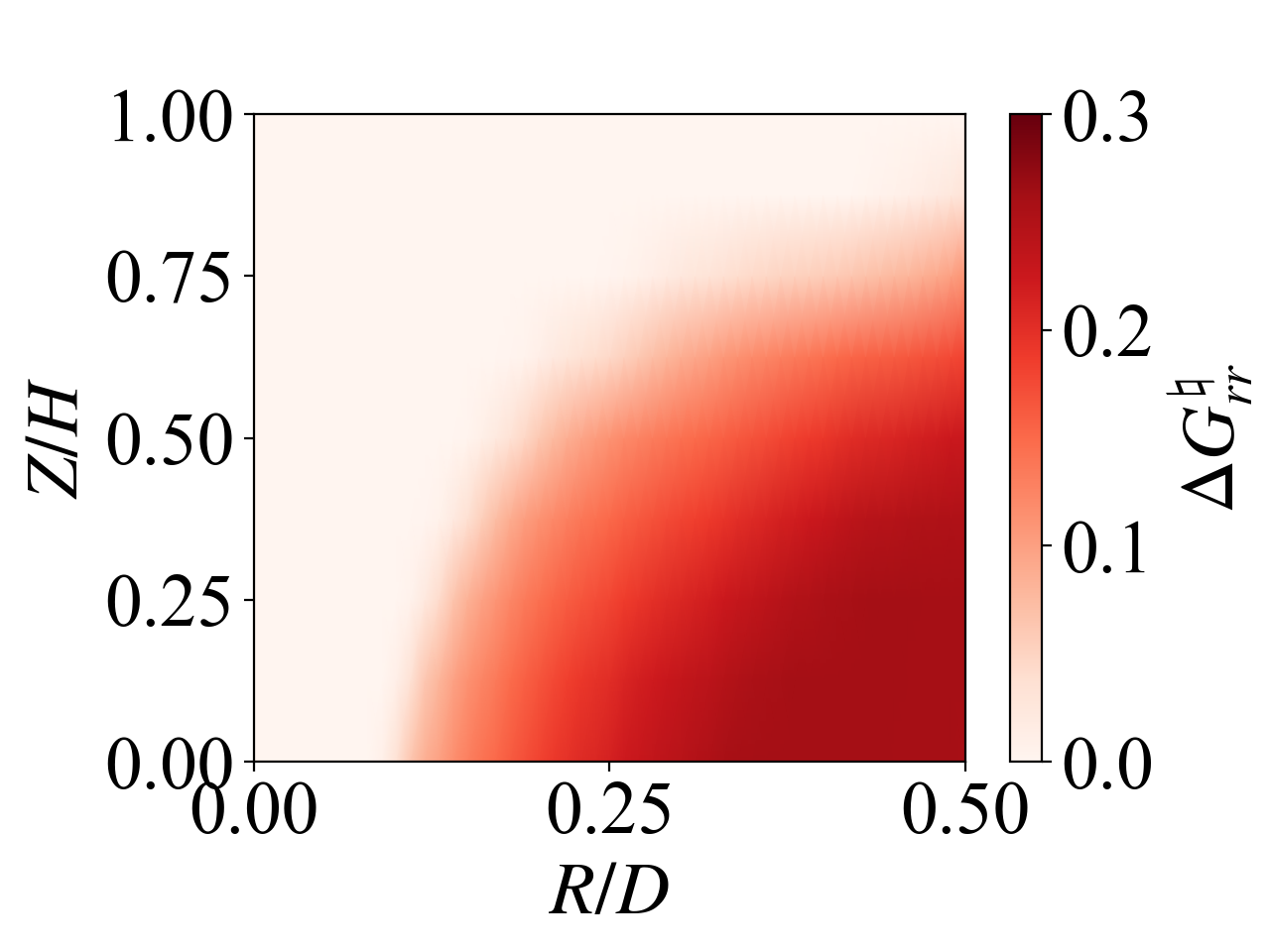}
		\caption{ 
			Spatial distribution of $\Delta G_{rr}^{\natural}$ at $\Theta=0$ after the rotation.
		}
		\label{fig:gr_RZ}
	\end{figure}

\begin{figure}[htbp]
		\centering
		\includegraphics[keepaspectratio, width=8cm]{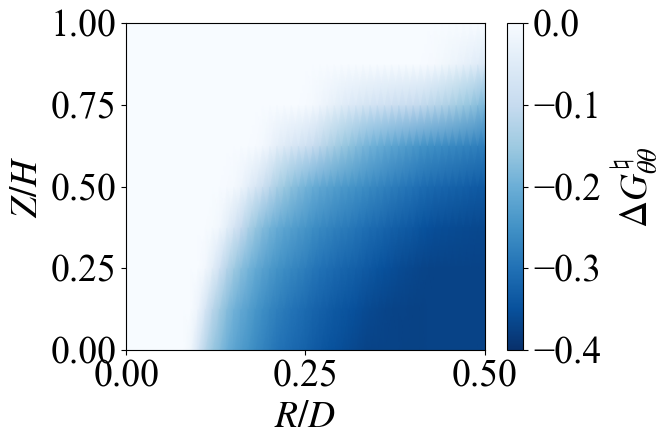}
		\caption{
			Spatial distribution of $\Delta G_{\theta \theta}^{\natural}$ at $\Theta=0$ after the rotation.
		}
		\label{fig:gt_RZ}
	\end{figure}

	Figures \ref{fig:gr_RZ} and \ref{fig:gt_RZ} show $\Delta G_{rr}^{\natural}$ and $\Delta G_{\theta \theta}^{\natural}$, respectively, as functions of $(R,Z)$ after the external oscillation.
	For the entire system, we find $\Delta G_{rr}^{\natural} \ge 0$ and $\Delta G_{\theta \theta}^{\natural} \le 0$.
	Near $2R/D=1$ and  $Z = 0$, $|\Delta G_{rr}^{\natural}|$ and $|\Delta G_{\theta \theta}^{\natural}|$ are largest.
	The spatial distributions are consistent with the residual stresses shown in Figs. \ref{fig:Sr_RZ} and \ref{fig:St_RZ}.
	These results indicate that the shear strain given by Eq. \eqref{Eq:deformation} causes the plastic deformation following Eqs. \eqref{Eq:evolve_Grr} and \eqref{Eq:evolve_Gtt}, resulting in the residual stress $S_{\theta \theta}>0$ consistent with the radial crack patterns in Fig. \ref{fig1}~(b).

	\section{Conclusions}
	\label{Sec:Summary}

	In this study, we numerically investigated the residual stress in a paste after rotating a cylindrical container based on an elasto-plastic model.
	The tension in the circumferential direction increases after the rotation as a memory effect.
	The residual stress is consistent with the experimentally observed radial crack pattern after desiccation \cite{Nakahara2005}.
	We analytically related the increase in tension to the change in the natural metric tensor characterizing plastic deformation.
	From the analysis of the time evolution of the natural metric tensor, we can theoretically explain the increase in tension in the circumferential direction.

\begin{figure}[htbp]
		\centering
		\includegraphics[keepaspectratio, width=8cm]{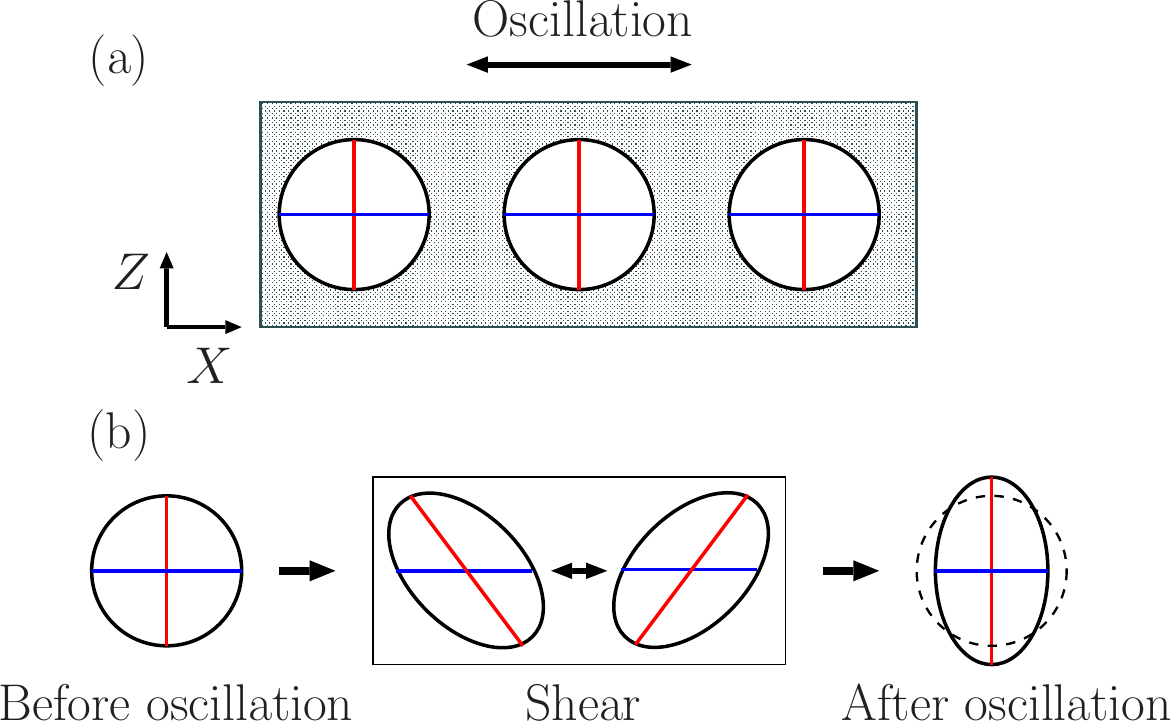}
		\caption{
			(a) Schematic illustration of a paste horizontally oscillated in one direction (the $X$-direction).
            The shapes of ellipses represent the plastic strain of the natural metric.
            (b) Schematic illustration of the nonlinear effect of shear motion due to the shear deformation due to the horizontal oscillation.
		}
		\label{fig:2D_model}
	\end{figure}

Equations \eqref{Eq:S_tt_remain} and \eqref{Eq:evolve_Gtt} describe the increase in tension after the rotation of the paste corresponding to Fig. \ref{fig1}(b).
The increase in tension after the oscillation in one direction corresponding to Fig. \ref{fig1}(a) is described by similar equations and explained by schematic illustrations in Fig. \ref{fig:2D_model} \cite{Ooshida2008,Ooshida2009,Goehring2015,Nakahara2019}.
        The plastic strain of the natural metric is represented by ellipses in Fig. \ref{fig:2D_model}(a).
        The deviatoric stress $S_{XX}$ is approximately given by
	\begin{align}
		\label{Eq:Sxx}
        S_{XX} \simeq -\mu \Delta g_{XX}^\natural
	\end{align}
    with $\Delta g_{XX}^\natural =g_{XX}^\natural - 1$, which is represented by the increase in the ellipse size in the $X$-direction.
    Here, $\Delta g_{XX}^\natural$ corresponds to the change in the natural length of a spring \cite{Ooshida2008,Ooshida2009}.
    The negative $\Delta g_{XX}^\natural$ indicates the shrinking of the ``natural length", which leads to the positive tension, $S_{XX}>0$, with Eq. \eqref{Eq:Sxx}.
    The time evolution of $\Delta g_{XX}^\natural$ is given by
	\begin{align}
		\label{Eq:evolve_Gxx}
		\partial_{t}\Delta g_{XX}^\natural &\simeq \tau^{-1}\left \{ -\frac{2}{3} \left ( \frac{du}{dz} \right )^{2} -\Delta g_{XX }^\natural \right \}
	\end{align}
    with the shear strain $\dfrac{du}{dz}$ in the $X$-$Z$ plane, where $\Delta g_{XX}^\natural$ becomes negative for non-zero $\dfrac{du}{dz}$ \cite{Ooshida2009}.
    Here, Eqs. \eqref{Eq:Sxx} and \eqref{Eq:evolve_Gxx} correspond to Eqs. \eqref{Eq:S_tt_remain} and \eqref{Eq:evolve_Gtt}, respectively.
    The evolution of $\Delta g_{XX}^\natural$ based on Eq. \eqref{Eq:evolve_Gxx} is intuitively explained in Fig. \ref{fig:2D_model}(b). In the initial state, the natural metric tensor is isotropic and represented by a circle.
    In the shear deformation due to the horizontal oscillation in the $X$-direction, the circle representing the natural metric is elongated along the $Z$-direction, while it shrinks in the $X$-direction due to the nonlinear effect shown in Eq. \eqref{Eq:evolve_Gxx}, which leads to the negative $\Delta g_{XX}^\natural$ and the residual tension $S_{XX}>0$ with Eq. \eqref{Eq:Sxx}.

\begin{figure}[htbp]
		\centering
		\includegraphics[keepaspectratio, width=7cm]{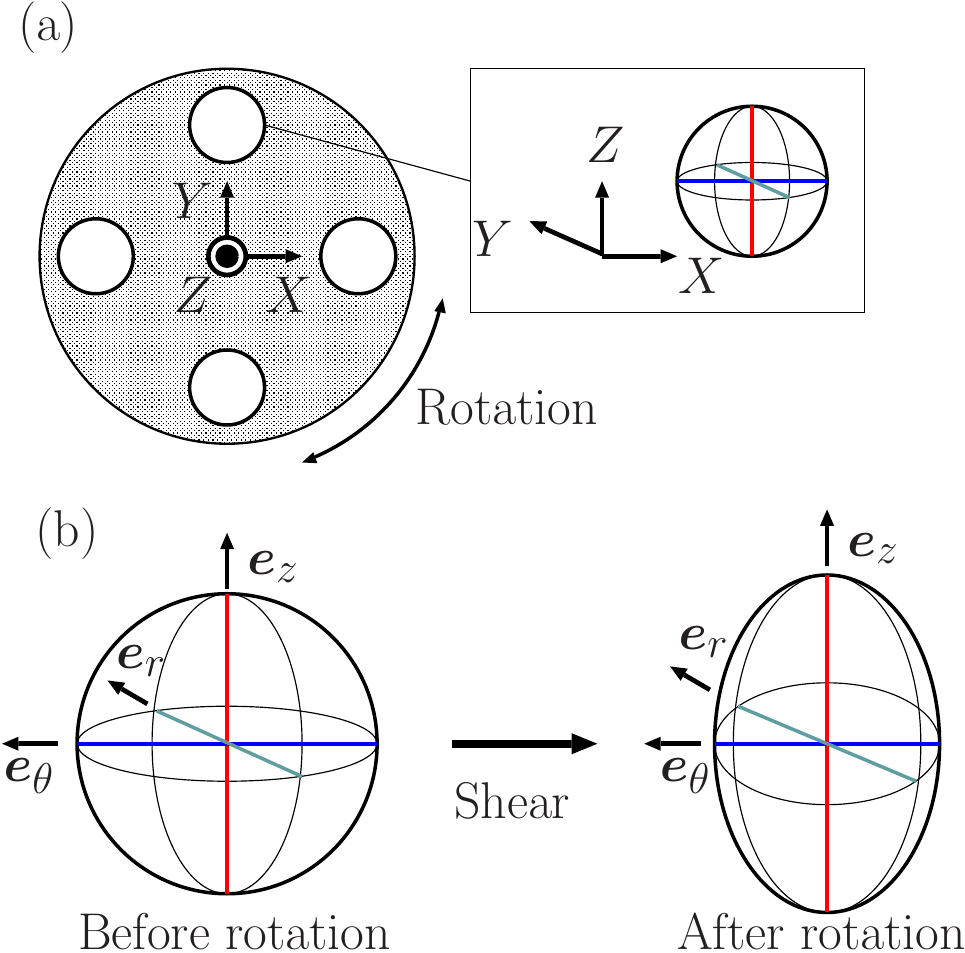}
		\caption{
			(a) Schematic illustration of a paste horizontally rotated.
            The shapes of ellipsoids represent the plastic strain of the natural metrick.
            (b) Schematic illustration of the nonlinear effect of shear deformation due to the horizontal rotation.
		}
		\label{fig:3D_model}
	\end{figure}

    The mechanism for the residual tension after rotation can be explained by a similar schematic illustration in Fig. \ref{fig:3D_model}.
    In Fig. \ref{fig:3D_model}(a), ellipsoids represent the plastic strain of the natural metric.
    The natural metric tensor is isotropic and represented by a sphere in the initial state in Fig. \ref{fig:3D_model}(b).
    After the shear strain in the $\theta$-$z$ plane due to horizontal rotation, the sphere is elongated along the $z$-direction, while it shrinks in the $\theta$-direction associated with $\Delta G_{\theta \theta}<0$, leading to residual tension $S_{\theta \theta}>0$ with Eq. \eqref{Eq:S_tt_remain}.
    Thus, the origins of the memory effect leading to the crack patterns in Fig. 1 (a) and (b) are essentially the same.
    
    For amorphous assemblies of particles, including dense colloidal suspensions (pastes), the relaxation time diverges due to the confinement of particles in cages formed by their neighbors when the volume fraction of particles is sufficiently high, which results in the non-zero yield stress \cite{Voigtmann2014}. 
    The residual stress induced by the rotation remains due to the divergence of the relaxation time during the drying process leading to the crack pattern in Fig. \ref{fig1}(a).
    In our theoretical model, we have introduced the constitutive equation with the yield stress phenomenologically, but we expect it to be derived from an extension of microscopic theories such as mode coupling theory \cite{Miyazaki2002,Fuchs2002,Ballauff2013,Mohan2013,Fritschi2014,Mohan2015,Moghimi2017} and pair distribution function theory\cite{Otsuki2006}.

	In our analysis, a large plastic deformation occurs near the wall of the cylindrical container, leading to maximum residual stress at the wall.
	Residual stress is enhanced by desiccation \cite{Kitsunezaki2016,Kitsunezaki2017}.
	Therefore, we expect that the radial crack in the experiments after desiccation is initiated at the wall boundary.
	However, further studies are required to confirm this.

	In experiments on pastes containing magnesium carbonate hydroxide \cite{Nakahara2006b}, radial crack patterns changed to ring patterns when a large deformation is applied.
	The ring crack patterns are expected to result from an increase in the residual tension in the radial direction.
	However, this residual stress distribution is not observed in our numerical simulations.
	An improvement in the elasto-plastic model will explain this behavior in the future.

	\begin{acknowledgments}
		The authors thank T. Ooshida, R. Tarumi, Y. Motoori, and S. Goto for the fruitful discussions.
		This study was partially supported by JSPS KAKENHI (Grant Nos. JP21H01006 ald JP23K03248).
		We would like to thank Editage (www.editage.jp) for the English language editing.
		%K.Y. appreciates using the supercomputer system of Research Center for Computational Science, Okazaki, Japan (Project: 22-IMS-C267 and 23-IMS-C126).
	\end{acknowledgments}

	\appendix

		\section{Approximation of left Cauchy-Green tensor}
		\label{App:B}
		
		In this appendix, we derive Eqs. \eqref{Eq:Br}, \eqref{Eq:Btheta}, and \eqref{Eq:Bz}.
		The left Cauchy-Green strain tensor $\bm{B}$ given by Eq. \eqref{Eq:B} is rewritten as
		\begin{align}
			\label{Eq:B_expand}
			\bm{B}=
			& g_{\natural}^{XX}\left(\frac{\partial \bm{r}}{\partial X}\otimes\frac{\partial \bm{r}}{\partial X}\right)
			+g_{\natural}^{XY} \left(\frac{\partial \bm{r}}{\partial X}\otimes\frac{\partial \bm{r}}{\partial Y}\right)\nonumber\\
			& +g_{\natural}^{XZ}\left(\frac{\partial \bm{r}}{\partial X}\otimes\frac{\partial \bm{r}}{\partial Z}\right)
			+g_{\natural}^{YX} \left(\frac{\partial \bm{r}}{\partial Y}\otimes\frac{\partial \bm{r}}{\partial X}\right)\nonumber\\
			& +g_{\natural}^{YY}\left(\frac{\partial \bm{r}}{\partial Y}\otimes\frac{\partial \bm{r}}{\partial Y}\right)
			+g_{\natural}^{YZ} \left(\frac{\partial \bm{r}}{\partial Y}\otimes\frac{\partial \bm{r}}{\partial Z}\right)\nonumber\\
			& +g_{\natural}^{ZX}\left(\frac{\partial \bm{r}}{\partial Z}\otimes\frac{\partial \bm{r}}{\partial X}\right)
			+g_{\natural}^{ZY} \left(\frac{\partial \bm{r}}{\partial Z}\otimes\frac{\partial \bm{r}}{\partial Y}\right)\nonumber\\
			& +g_{\natural}^{ZZ}\left(\frac{\partial \bm{r}}{\partial Z}\otimes\frac{\partial \bm{r}}{\partial Z}\right).
		\end{align}
		Here, $\dfrac{\partial \bm{r}}{\partial X_i}$ is written as
		\begin{align}
			\label{Eq:r/X}
			\frac{\partial \bm{r}}{\partial X_{i}}=\frac{\partial R}{\partial X_{i}}\frac{\partial \bm{r}}{\partial R}+\frac{\partial \Theta}{\partial X_{i}}\frac{\partial \bm{r}}{\partial \Theta}+\frac{\partial Z}{\partial X_{i}}\frac{\partial \bm{r}}{\partial Z}.
		\end{align}
		Equation \eqref{Eq:R_Theta} gives
		\begin{align}
			\label{Eq:partial_R}
			\frac{\partial R}{\partial X} &=\cos\Theta  &\frac{\partial R}{\partial Y} &=\sin\Theta &\frac{\partial R}{\partial Z}&=0,\\
			\label{Eq:partial_Theta}
			\frac{\partial \Theta}{\partial X} &=-\frac{\sin\Theta}{R}  &\frac{\partial \Theta}{\partial Y} &=\frac{\cos\Theta}{R} &\frac{\partial \Theta}{\partial Z}&=0\\
			\label{Eq:partial_Z}
			\frac{\partial Z}{\partial X} &=0  &\frac{\partial Z}{\partial Y} &=0 &\frac{\partial Z}{\partial Z}&=1.
		\end{align}
		From Eq. \eqref{Eq:deformation}, we obtain
		\begin{align}
			\label{Eq:partial_r_cy}
			\frac{\partial \bm{r}}{\partial R}=\bm{e}_{r}, \quad
			\frac{\partial \bm{r}}{\partial \Theta}=R\bm{e}_{\theta}, \quad
			\frac{\partial \bm{r}}{\partial Z}=R\left ( \partial_Z \varphi \right ) \bm{e}_{\theta}+\bm{e}_{z}.
		\end{align}
		Substituting Eqs. \eqref{Eq:partial_R}-\eqref{Eq:partial_r_cy} into Eq. \eqref{Eq:r/X}, we obtain 
		\begin{align}
			\label{Eq:partial_r_X}
			\frac{\partial \bm{r}}{\partial X}&=\cos\Theta\bm{e}_{r} -\sin\Theta\bm{e}_{\theta}, \\
			\label{Eq:partial_r_Y}
			\frac{\partial \bm{r}}{\partial Y}&=\sin\Theta\bm{e}_{r} +\cos\Theta\bm{e}_{\theta},\\
			\label{Eq:partial_r_Z}
			\frac{\partial \bm{r}}{\partial Z}&=R\left (\partial_{Z}\varphi \right )\bm{e}_{\theta}+\bm{e}_{z}.
		\end{align}

		The inverse matrix $\bm{g}_\natural$ in Eq. \eqref{Eq:B_expand} is represented as
		\begin{align}
			\label{Eq:gnat_G}
			\bm{g}_\natural = \bm{Q}^{\rm T} \bm{G}_\natural \bm{Q},
		\end{align}
		where $\bm{G}_\natural$ satisfies
		\begin{align}
			\label{Eq:G_natural_inv}
			\bm{G}_\natural \bm{G}^\natural = \openone.
		\end{align}
		Considering a small deformation of $\varphi \ll 1$, $\bm{G}_\natural$ can be approximated as
		\begin{align}
			\label{Eq:G_natural_inv_app}
			\bm{G}_\natural \simeq \left(
			\begin{array}{ccc}
				1-\Delta G_{rr}^\natural & -\gamma & -\delta \\
				-\gamma&1-\Delta G_{\theta \theta}^\natural& -\epsilon \\
				-\delta & -\epsilon & 1 + \Delta G_{rr}^\natural + \Delta G_{\theta \theta}^\natural
			\end{array}
			\right ).
		\end{align}
		Substituting Eq. \eqref{Eq:G_natural_inv_app} into Eq. \eqref{Eq:gnat_G}, $(g_\natural^{ij})$ is written as
		\begin{align}
			g_\natural^{XX} = & (1-\Delta G_{rr}^\natural) \cos^2 \theta +(1-\Delta G_{\theta \theta}^\natural) \sin^2 \theta \nonumber \\
			& +2 \gamma \sin \theta \cos \theta, \\
			g_\natural^{YY} = & (1-\Delta G_{rr}^\natural) \sin^2 \theta +(1-\Delta G_{\theta \theta}^\natural) \cos^2 \theta \nonumber \\
			& -2 \gamma \sin \theta \cos \theta, \\
			g_\natural^{ZZ} = & 1+\Delta G_{rr}^\natural +\Delta G_{\theta \theta}^\natural, \\
			g_\natural^{XY} = & g_\natural^{YX} = (\Delta G_{\theta \theta}^\natural-\Delta G_{rr}^\natural)\sin \theta \cos \theta \nonumber \\
			& - \gamma (\cos^2 \theta - \sin^2 \theta), \\
			g_\natural^{YZ} = & g_\natural^{ZY} =  - \delta \sin \theta - \epsilon \cos \theta, \\
			g_\natural^{ZX} = & g_\natural^{XZ} =  - \delta \cos \theta + \epsilon \sin \theta.
		\end{align}

		Substituting Eq. \eqref{Eq:B_expand} with these equations and Eqs. \eqref{Eq:partial_r_X}-\eqref{Eq:partial_r_Z} into Eq. \eqref{Eq:B_cy}, we obtain
		\begin{align}
			\label{Eq:Br_G}
			B_{rr}  = & (1- \Delta G^\natural_{rr}) \cos^2 \varphi + (1 - \Delta G^\natural_{\theta \theta}) \sin^2 \varphi \nonumber \\
			&+2 \gamma \sin \varphi \cos \varphi,\\
			\label{Eq:Btheta_G}
			B_{\theta \theta}  = & (1- \Delta G^\natural_{rr}) \sin^2 \varphi + (1 - \Delta G^\natural_{\theta \theta}) \cos^2 \varphi \nonumber \\
			& -2 \gamma \sin \varphi \cos \varphi  \nonumber \\
			& - 2 \left (\delta \sin \varphi + \epsilon \cos \varphi  \right ) R \left ( \partial_Z \varphi \right ) \nonumber \\
			& +  (1+\Delta G_{rr}^\natural +\Delta G_{\theta \theta}^\natural) R^2\left ( \partial_Z \varphi \right )^2,  \\
			\label{Eq:Bz_G}
			B_{zz}  = & 1+\Delta G_{rr}^\natural+\Delta G_{\theta \theta}^\natural.
		\end{align}
		Neglecting the higher-order terms of $\Delta G^\natural_{rr}$, $\Delta G^\natural_{\theta \theta}$, $\gamma$, $\delta$, $\epsilon$, and $\varphi$ from these euqations, we obtain Eqs. \eqref{Eq:Br}, \eqref{Eq:Btheta}, and \eqref{Eq:Bz}.

		\section{Time evolution of the natural metric tensor}
		\label{App:DG}
		
		In this appendix, we derive Eqs. \eqref{Eq:evolve_Grr} and \eqref{Eq:evolve_Gtt}.
		From Eq. \eqref{Eq:gnat_G}, $\bm{G}_\natural$ is represented as
		\begin{align}
			\bm{G}_{\natural}= \bm{Q} \bm{g}_{\natural} \bm{Q}^{\mathrm{T}}.
		\end{align}
		By differentiating this equation by $t$, we obtain
		\begin{align}
			\label{Eq:der_G}
			\partial_{t} \bm{G}_{\natural}= \bm{Q} \left ( \partial _t \bm{g}_\natural \right ) \bm{Q}^{\mathrm{T}} + \left (\partial _t\bm{Q} \right )\bm{Q}^{\mathrm{T}} \bm{G}_\natural  + \left \{ \left (\partial _t\bm{Q} \right )\bm{Q}^{\mathrm{T}} \bm{G}_\natural  \right \}^{\rm T}.
		\end{align}
		Here, we introduce 
		\begin{align}
			\label{Eq:G_def}
			\bm{G} =  \bm{Q} \bm{g}  \bm{Q}^{\mathrm{T}}
		\end{align}
		and its inverse
		\begin{align}
			\label{Eq:G_inv}
			\bm{G}^{-1} =  \bm{Q} \bm{g}^{-1}  \bm{Q}^{\mathrm{T}}.
		\end{align}
		From Eq. \eqref{Eq:evolve_g}, we obtain
		\begin{align}
			\label{Eq:der_QGQ}
			\bm{Q} \left ( \partial _t \bm{g}_\natural \right ) \bm{Q}^{\mathrm{T}} = \tau^{-1} \left ( {\Gamma}\bm{G}^{-1}-\bm{G}_{\natural} \right ).
		\end{align}
		Using Eqs. \eqref{Eq:Q} and \eqref{Eq:G_natural_inv_app}, the second and third terms on the right-hand side of Eq. \eqref{Eq:der_G} are written as
		\begin{align}
			\label{Eq:dQQG}
			\left (\partial _t\bm{Q} \right )\bm{Q}^{\mathrm{T}} \bm{G}_\natural  + \left \{ \left (\partial _t\bm{Q} \right )\bm{Q}^{\mathrm{T}} \bm{G}_\natural  \right \}^{\rm T} = \tilde{\bm{G}}_\natural \partial_t \varphi
		\end{align}
		with\begin{align}
			\label{Eq:tG}
			\tilde{\bm{G}}_\natural =
			\left(
			\begin{array}{ccc}
				-2 \gamma & \Delta G_{rr}^\natural - \Delta G_{\theta \theta}^\natural & - \epsilon\\
				\Delta G_{rr}^\natural - \Delta G_{\theta \theta}^\natural &2 \gamma& \delta \\
				-\epsilon & \delta & 0
			\end{array}
			\right ).
		\end{align}
		Substituting Eqs. \eqref{Eq:der_QGQ} and \eqref{Eq:dQQG} into Eq. \eqref{Eq:der_G}, we obtain
		\begin{align}
			\label{Eq:evolve_G_vec}
			\partial_{t} \bm{G}_{\natural} - \tilde{\bm{G}}_{\natural} \partial_t \varphi = \tau^{-1} \left ( {\Gamma}\bm{G}^{-1}-\bm{G}_{\natural} \right ).
		\end{align}
		
		From Eqs. \eqref{Eq:partial_r_X}, \eqref{Eq:partial_r_Y}, and \eqref{Eq:partial_r_Z} with Eq. \eqref{Eq:gij}, the natural metric tensor is given by
		\begin{align}
			\label{Eq:g}
			\bm{g} & =
			\left(
			\begin{array}{ccc}
				g_{XX} & g_{XY}  & g_{XZ}   \\
				g_{YX} & g_{YY}  & g_{YZ}   \\
				g_{ZX} & g_{ZY}  & g_{ZZ}  
			\end{array}
			\right ) \nonumber \\
			& =
			\left(
			\begin{array}{ccc}
				1 & 0 &  -R \left ( \partial_Z \varphi \right ) \sin \Theta \\
				0&1&  R \left ( \partial_Z \varphi \right ) \cos \Theta  \\
				-R \left ( \partial_Z \varphi \right ) \sin \Theta& R \left ( \partial_Z \varphi \right ) \cos \Theta  & 1 + R^2 \left ( \partial_Z \varphi \right )^2
			\end{array}
			\right ).
		\end{align}
		Substituting Eq. \eqref{Eq:g} into Eq. \eqref{Eq:G_def} with Eq. \eqref{Eq:Q}, we obtain
		\begin{align}
			\label{Eq:G}
			\bm{G} =
			\left(
			\begin{array}{ccc}
				1 & 0 &  R \left ( \partial_Z \varphi \right ) \sin \varphi \\
				0&1&  R \left ( \partial_Z \varphi \right ) \cos \varphi  \\
				R \left ( \partial_Z \varphi \right ) \sin \varphi& R \left ( \partial_Z \varphi \right ) \cos \varphi  & 1 + R^2 \left ( \partial_Z \varphi \right )^2
			\end{array}
			\right ).
		\end{align}
		The inverse matrix $\bm{G}^{-1}$ is approximately obtained from Eq. \eqref{Eq:G} as
		\begin{align}
			\label{Eq:G_inv_app}
			\bm{G}^{-1}\simeq 
			\left(
			\begin{array}{ccc}
				1  & 0 &  -R\varphi \left ( \partial_Z \varphi \right )  \\
				0 &1+ R^2 \left ( \partial_Z \varphi \right )^2 &  -R \left ( \partial_Z \varphi \right )  \\
				-R \varphi \left ( \partial_Z \varphi \right )& -R \left ( \partial_Z \varphi \right )  & 1 
			\end{array}
			\right ).
		\end{align}

		The coefficient $\Gamma$ given by Eq. \eqref{Eq:Gamma} can be expressed as follows:
		\begin{eqnarray}
			\label{Eq:Gamma_tr}
			\Gamma=\frac{3}{\mathrm{ tr} \left ( \bm{g}^{-1}\bm{g}^{\natural} \right )} = \frac{3}{\mathrm{ tr} \left ( \bm{G}^{-1}\bm{G}^{\natural} \right )}.
		\end{eqnarray}
		Substituting Eqs. \eqref{Eq:G_natural_app} and \eqref{Eq:G_inv_app} into Eq. \eqref{Eq:Gamma_tr}, $\Gamma$ is approximately represented as
		\begin{eqnarray}
			\label{Eq:Gamma_app}
			\Gamma \simeq 1 - \frac{1}{3}R^2 \left ( \partial_Z \varphi \right )^2.
		\end{eqnarray}
		Using Eqs. \eqref{Eq:G_inv_app} and \eqref{Eq:Gamma_app}, we obtain
		\begin{align}
			\label{Eq:GG_inv}
			\Gamma \bm{G}^{-1}\simeq 
			\left(
			\begin{array}{ccc}
				1 - \frac{1}{3}R^2 \left ( \partial_Z \varphi \right )^2& 0 &  -R\varphi \left ( \partial_Z \varphi \right )  \\
				0 & 1 + \frac{2}{3}R^2 \left ( \partial_Z \varphi \right )^2 &  -R \left ( \partial_Z \varphi \right )  \\
				-R \varphi \left ( \partial_Z \varphi \right )& -R \left ( \partial_Z \varphi \right )  & 1  - \frac{1}{3}R^2 \left ( \partial_Z \varphi \right )^2
			\end{array}
			\right ).
		\end{align}
		
		Substituting Eqs. \eqref{Eq:G_natural_app}, \eqref{Eq:tG}, \eqref{Eq:GG_inv} into Eq. \eqref{Eq:evolve_G_vec}, we derive the time evolution equations for $\Delta G_{rr}$ and $\Delta G_{\theta \theta}$ as Eqs. \eqref{Eq:evolve_Grr} and \eqref{Eq:evolve_Gtt}.

		\bibliography{reference}% Produces the bibliography via BibTeX.

\begin{thebibliography}{52}%
\makeatletter
\providecommand \@ifxundefined [1]{%
 \@ifx{#1\undefined}
}%
\providecommand \@ifnum [1]{%
 \ifnum #1\expandafter \@firstoftwo
 \else \expandafter \@secondoftwo
 \fi
}%
\providecommand \@ifx [1]{%
 \ifx #1\expandafter \@firstoftwo
 \else \expandafter \@secondoftwo
 \fi
}%
\providecommand \natexlab [1]{#1}%
\providecommand \enquote  [1]{``#1''}%
\providecommand \bibnamefont  [1]{#1}%
\providecommand \bibfnamefont [1]{#1}%
\providecommand \citenamefont [1]{#1}%
\providecommand \href@noop [0]{\@secondoftwo}%
\providecommand \href [0]{\begingroup \@sanitize@url \@href}%
\providecommand \@href[1]{\@@startlink{#1}\@@href}%
\providecommand \@@href[1]{\endgroup#1\@@endlink}%
\providecommand \@sanitize@url [0]{\catcode `\\12\catcode `\$12\catcode
  `\&12\catcode `\#12\catcode `\^12\catcode `\_12\catcode `\%12\relax}%
\providecommand \@@startlink[1]{}%
\providecommand \@@endlink[0]{}%
\providecommand \url  [0]{\begingroup\@sanitize@url \@url }%
\providecommand \@url [1]{\endgroup\@href {#1}{\urlprefix }}%
\providecommand \urlprefix  [0]{URL }%
\providecommand \Eprint [0]{\href }%
\providecommand \doibase [0]{https://doi.org/}%
\providecommand \selectlanguage [0]{\@gobble}%
\providecommand \bibinfo  [0]{\@secondoftwo}%
\providecommand \bibfield  [0]{\@secondoftwo}%
\providecommand \translation [1]{[#1]}%
\providecommand \BibitemOpen [0]{}%
\providecommand \bibitemStop [0]{}%
\providecommand \bibitemNoStop [0]{.\EOS\space}%
\providecommand \EOS [0]{\spacefactor3000\relax}%
\providecommand \BibitemShut  [1]{\csname bibitem#1\endcsname}%
\let\auto@bib@innerbib\@empty
%</preamble>
\bibitem [{\citenamefont {Goehring}\ \emph {et~al.}(2015)\citenamefont
  {Goehring}, \citenamefont {Nakahara}, \citenamefont {Dutta}, \citenamefont
  {Kitsunezaki},\ and\ \citenamefont {Tarafdar}}]{Goehring2015}%
  \BibitemOpen
  \bibfield  {author} {\bibinfo {author} {\bibfnamefont {L.}~\bibnamefont
  {Goehring}}, \bibinfo {author} {\bibfnamefont {A.}~\bibnamefont {Nakahara}},
  \bibinfo {author} {\bibfnamefont {T.}~\bibnamefont {Dutta}}, \bibinfo
  {author} {\bibfnamefont {S.}~\bibnamefont {Kitsunezaki}},\ and\ \bibinfo
  {author} {\bibfnamefont {S.}~\bibnamefont {Tarafdar}},\ }\href
  {https://doi.org/10.1002/9783527671922} {\emph {\bibinfo {title} {Desiccation
  Cracks and their Patterns}}}\ (\bibinfo  {publisher} {Wiley},\ \bibinfo
  {year} {2015})\BibitemShut {NoStop}%
\bibitem [{\citenamefont {Kindle}(1917)}]{Kindle1917}%
  \BibitemOpen
  \bibfield  {author} {\bibinfo {author} {\bibfnamefont {E.~M.}\ \bibnamefont
  {Kindle}},\ }\bibfield  {title} {\bibinfo {title} {Some factors affecting the
  development of mud-cracks},\ }\href {https://doi.org/10.1086/622446}
  {\bibfield  {journal} {\bibinfo  {journal} {J. Geol.}\ }\textbf {\bibinfo
  {volume} {25}},\ \bibinfo {pages} {135} (\bibinfo {year} {1917})}\BibitemShut
  {NoStop}%
\bibitem [{\citenamefont {Groisman}\ and\ \citenamefont
  {Kaplan}(1994)}]{Groisman1994}%
  \BibitemOpen
  \bibfield  {author} {\bibinfo {author} {\bibfnamefont {A.}~\bibnamefont
  {Groisman}}\ and\ \bibinfo {author} {\bibfnamefont {E.}~\bibnamefont
  {Kaplan}},\ }\bibfield  {title} {\bibinfo {title} {An experimental study of
  cracking induced by desiccation},\ }\href
  {https://doi.org/10.1209/0295-5075/25/6/004} {\bibfield  {journal} {\bibinfo
  {journal} {EPL}\ }\textbf {\bibinfo {volume} {25}},\ \bibinfo {pages} {415}
  (\bibinfo {year} {1994})}\BibitemShut {NoStop}%
\bibitem [{\citenamefont {Bohn}\ \emph {et~al.}(2005)\citenamefont {Bohn},
  \citenamefont {Pauchard},\ and\ \citenamefont {Couder}}]{Bohn2005}%
  \BibitemOpen
  \bibfield  {author} {\bibinfo {author} {\bibfnamefont {S.}~\bibnamefont
  {Bohn}}, \bibinfo {author} {\bibfnamefont {L.}~\bibnamefont {Pauchard}},\
  and\ \bibinfo {author} {\bibfnamefont {Y.}~\bibnamefont {Couder}},\
  }\bibfield  {title} {\bibinfo {title} {Hierarchical crack pattern as formed
  by successive domain divisions.},\ }\href
  {https://doi.org/10.1103/PhysRevE.71.046214} {\bibfield  {journal} {\bibinfo
  {journal} {Phys. Rev. E}\ }\textbf {\bibinfo {volume} {71}},\ \bibinfo
  {pages} {046214} (\bibinfo {year} {2005})}\BibitemShut {NoStop}%
\bibitem [{\citenamefont {Shorlin}\ \emph {et~al.}(2000)\citenamefont
  {Shorlin}, \citenamefont {de~Bruyn}, \citenamefont {Graham},\ and\
  \citenamefont {Morris}}]{Shorlin2000}%
  \BibitemOpen
  \bibfield  {author} {\bibinfo {author} {\bibfnamefont {K.~A.}\ \bibnamefont
  {Shorlin}}, \bibinfo {author} {\bibfnamefont {J.~R.}\ \bibnamefont
  {de~Bruyn}}, \bibinfo {author} {\bibfnamefont {M.}~\bibnamefont {Graham}},\
  and\ \bibinfo {author} {\bibfnamefont {S.~W.}\ \bibnamefont {Morris}},\
  }\bibfield  {title} {\bibinfo {title} {Development and geometry of isotropic
  and directional shrinkage-crack patterns},\ }\href
  {https://doi.org/10.1103/PhysRevE.61.6950} {\bibfield  {journal} {\bibinfo
  {journal} {Phys. Rev. E}\ }\textbf {\bibinfo {volume} {61}},\ \bibinfo
  {pages} {6950} (\bibinfo {year} {2000})}\BibitemShut {NoStop}%
\bibitem [{\citenamefont {Goehring}\ \emph {et~al.}(2010)\citenamefont
  {Goehring}, \citenamefont {Conroy}, \citenamefont {Akhter}, \citenamefont
  {Clegg},\ and\ \citenamefont {Routh}}]{Goehring2010}%
  \BibitemOpen
  \bibfield  {author} {\bibinfo {author} {\bibfnamefont {L.}~\bibnamefont
  {Goehring}}, \bibinfo {author} {\bibfnamefont {R.}~\bibnamefont {Conroy}},
  \bibinfo {author} {\bibfnamefont {A.}~\bibnamefont {Akhter}}, \bibinfo
  {author} {\bibfnamefont {W.~J.}\ \bibnamefont {Clegg}},\ and\ \bibinfo
  {author} {\bibfnamefont {A.~F.}\ \bibnamefont {Routh}},\ }\bibfield  {title}
  {\bibinfo {title} {Evolution of mud-crack patterns during repeated drying
  cycles},\ }\href {https://doi.org/10.1039/b922206e} {\bibfield  {journal}
  {\bibinfo  {journal} {Soft Matter}\ }\textbf {\bibinfo {volume} {6}},\
  \bibinfo {pages} {3562} (\bibinfo {year} {2010})}\BibitemShut {NoStop}%
\bibitem [{\citenamefont {Nakahara}\ and\ \citenamefont
  {Matsuo}(2005)}]{Nakahara2005}%
  \BibitemOpen
  \bibfield  {author} {\bibinfo {author} {\bibfnamefont {A.}~\bibnamefont
  {Nakahara}}\ and\ \bibinfo {author} {\bibfnamefont {Y.}~\bibnamefont
  {Matsuo}},\ }\bibfield  {title} {\bibinfo {title} {Imprinting memory into
  paste and its visualization as crack patterns in drying process},\ }\href
  {https://doi.org/10.1143/JPSJ.74.1362} {\bibfield  {journal} {\bibinfo
  {journal} {J. Phys. Soc. Jpn.}\ }\textbf {\bibinfo {volume} {74}},\ \bibinfo
  {pages} {1362} (\bibinfo {year} {2005})}\BibitemShut {NoStop}%
\bibitem [{\citenamefont {Nakahara}\ and\ \citenamefont
  {Matsuo}(2006{\natexlab{a}})}]{Nakahara2006a}%
  \BibitemOpen
  \bibfield  {author} {\bibinfo {author} {\bibfnamefont {A.}~\bibnamefont
  {Nakahara}}\ and\ \bibinfo {author} {\bibfnamefont {Y.}~\bibnamefont
  {Matsuo}},\ }\bibfield  {title} {\bibinfo {title} {Imprinting memory into
  paste to control crack formation in drying process},\ }\href
  {https://doi.org/10.1088/1742-5468/2006/07/P07016} {\bibfield  {journal}
  {\bibinfo  {journal} {J. Stat. Mech: Theory Exp.}\ }\textbf {\bibinfo
  {volume} {2006}},\ \bibinfo {pages} {P07016} (\bibinfo {year}
  {2006}{\natexlab{a}})}\BibitemShut {NoStop}%
\bibitem [{\citenamefont {Nakahara}\ and\ \citenamefont
  {Matsuo}(2006{\natexlab{b}})}]{Nakahara2006b}%
  \BibitemOpen
  \bibfield  {author} {\bibinfo {author} {\bibfnamefont {A.}~\bibnamefont
  {Nakahara}}\ and\ \bibinfo {author} {\bibfnamefont {Y.}~\bibnamefont
  {Matsuo}},\ }\bibfield  {title} {\bibinfo {title} {Transition in the pattern
  of cracks resulting from memory effects in paste},\ }\href
  {https://doi.org/10.1103/PhysRevE.74.045102} {\bibfield  {journal} {\bibinfo
  {journal} {Phys. Rev. E}\ }\textbf {\bibinfo {volume} {74}},\ \bibinfo
  {pages} {045102} (\bibinfo {year} {2006}{\natexlab{b}})}\BibitemShut
  {NoStop}%
\bibitem [{\citenamefont {Matsuo}\ and\ \citenamefont
  {Nakahara}(2012)}]{Matsuo2012}%
  \BibitemOpen
  \bibfield  {author} {\bibinfo {author} {\bibfnamefont {Y.}~\bibnamefont
  {Matsuo}}\ and\ \bibinfo {author} {\bibfnamefont {A.}~\bibnamefont
  {Nakahara}},\ }\bibfield  {title} {\bibinfo {title} {Effect of interaction on
  the formation of memories in paste},\ }\href
  {https://doi.org/10.1143/JPSJ.81.024801} {\bibfield  {journal} {\bibinfo
  {journal} {J. Phys. Soc. Jpn.}\ }\textbf {\bibinfo {volume} {81}},\ \bibinfo
  {pages} {024801} (\bibinfo {year} {2012})}\BibitemShut {NoStop}%
\bibitem [{\citenamefont {Nakayama}\ \emph {et~al.}(2013)\citenamefont
  {Nakayama}, \citenamefont {Matsuo}, \citenamefont {Takeshi},\ and\
  \citenamefont {Nakahara}}]{Nakayama2013}%
  \BibitemOpen
  \bibfield  {author} {\bibinfo {author} {\bibfnamefont {H.}~\bibnamefont
  {Nakayama}}, \bibinfo {author} {\bibfnamefont {Y.}~\bibnamefont {Matsuo}},
  \bibinfo {author} {\bibfnamefont {O.}~\bibnamefont {Takeshi}},\ and\ \bibinfo
  {author} {\bibfnamefont {A.}~\bibnamefont {Nakahara}},\ }\bibfield  {title}
  {\bibinfo {title} {Position control of desiccation cracks by memory effect
  and faraday waves},\ }\href {https://doi.org/10.1140/epje/i2013-13001-8}
  {\bibfield  {journal} {\bibinfo  {journal} {Eur. Phys. J. E}\ }\textbf
  {\bibinfo {volume} {36}},\ \bibinfo {pages} {1} (\bibinfo {year}
  {2013})}\BibitemShut {NoStop}%
\bibitem [{\citenamefont {Mal}\ \emph {et~al.}(2005)\citenamefont {Mal},
  \citenamefont {Sinha}, \citenamefont {Mitra},\ and\ \citenamefont
  {Tarafdar}}]{Mal2005}%
  \BibitemOpen
  \bibfield  {author} {\bibinfo {author} {\bibfnamefont {D.}~\bibnamefont
  {Mal}}, \bibinfo {author} {\bibfnamefont {S.}~\bibnamefont {Sinha}}, \bibinfo
  {author} {\bibfnamefont {S.}~\bibnamefont {Mitra}},\ and\ \bibinfo {author}
  {\bibfnamefont {S.}~\bibnamefont {Tarafdar}},\ }\bibfield  {title} {\bibinfo
  {title} {Formation of crack networks in drying laponite films},\ }\href
  {https://doi.org/10.1016/j.physa.2004.08.056} {\bibfield  {journal} {\bibinfo
   {journal} {Physica A}\ }\textbf {\bibinfo {volume} {346}},\ \bibinfo {pages}
  {110} (\bibinfo {year} {2005})}\BibitemShut {NoStop}%
\bibitem [{\citenamefont {Khatun}\ \emph {et~al.}(2013)\citenamefont {Khatun},
  \citenamefont {Dutta},\ and\ \citenamefont {Tarafdar}}]{Khatun2013}%
  \BibitemOpen
  \bibfield  {author} {\bibinfo {author} {\bibfnamefont {T.}~\bibnamefont
  {Khatun}}, \bibinfo {author} {\bibfnamefont {T.}~\bibnamefont {Dutta}},\ and\
  \bibinfo {author} {\bibfnamefont {S.}~\bibnamefont {Tarafdar}},\ }\bibfield
  {title} {\bibinfo {title} {Crack formation under an electric field in
  droplets of laponite gel: Memory effect and scaling relations},\ }\href
  {https://doi.org/10.1021/la404297k} {\bibfield  {journal} {\bibinfo
  {journal} {Langmuir}\ }\textbf {\bibinfo {volume} {29}},\ \bibinfo {pages}
  {15535} (\bibinfo {year} {2013})}\BibitemShut {NoStop}%
\bibitem [{\citenamefont {Pauchard}\ \emph {et~al.}(2008)\citenamefont
  {Pauchard}, \citenamefont {Elias}, \citenamefont {Boltenhagen}, \citenamefont
  {Cebers},\ and\ \citenamefont {Bacri}}]{Pauchard2008}%
  \BibitemOpen
  \bibfield  {author} {\bibinfo {author} {\bibfnamefont {L.}~\bibnamefont
  {Pauchard}}, \bibinfo {author} {\bibfnamefont {F.}~\bibnamefont {Elias}},
  \bibinfo {author} {\bibfnamefont {P.}~\bibnamefont {Boltenhagen}}, \bibinfo
  {author} {\bibfnamefont {A.}~\bibnamefont {Cebers}},\ and\ \bibinfo {author}
  {\bibfnamefont {J.~C.}\ \bibnamefont {Bacri}},\ }\bibfield  {title} {\bibinfo
  {title} {When a crack is oriented by a magnetic field},\ }\href
  {https://doi.org/10.1103/PhysRevE.77.021402} {\bibfield  {journal} {\bibinfo
  {journal} {Phys. Rev. E}\ }\textbf {\bibinfo {volume} {77}},\ \bibinfo
  {pages} {021402} (\bibinfo {year} {2008})}\BibitemShut {NoStop}%
\bibitem [{\citenamefont {Ngo}\ \emph {et~al.}(2008)\citenamefont {Ngo},
  \citenamefont {Richardi},\ and\ \citenamefont {Pileni}}]{Ngo2008}%
  \BibitemOpen
  \bibfield  {author} {\bibinfo {author} {\bibfnamefont {A.~T.}\ \bibnamefont
  {Ngo}}, \bibinfo {author} {\bibfnamefont {J.}~\bibnamefont {Richardi}},\ and\
  \bibinfo {author} {\bibfnamefont {M.~P.}\ \bibnamefont {Pileni}},\ }\bibfield
   {title} {\bibinfo {title} {Do directional primary and secondary crack
  patterns in thin films of maghemite nanocrystals follow a universal scaling
  law?},\ }\href {https://doi.org/10.1021/jp802736g} {\bibfield  {journal}
  {\bibinfo  {journal} {J. Phys. Chem. B}\ }\textbf {\bibinfo {volume} {112}},\
  \bibinfo {pages} {14409} (\bibinfo {year} {2008})}\BibitemShut {NoStop}%
\bibitem [{\citenamefont {Lama}\ \emph {et~al.}(2016)\citenamefont {Lama},
  \citenamefont {Dugyala}, \citenamefont {Basavaraj},\ and\ \citenamefont
  {Satapathy}}]{Lama2016}%
  \BibitemOpen
  \bibfield  {author} {\bibinfo {author} {\bibfnamefont {H.}~\bibnamefont
  {Lama}}, \bibinfo {author} {\bibfnamefont {V.~R.}\ \bibnamefont {Dugyala}},
  \bibinfo {author} {\bibfnamefont {M.~G.}\ \bibnamefont {Basavaraj}},\ and\
  \bibinfo {author} {\bibfnamefont {D.~K.}\ \bibnamefont {Satapathy}},\
  }\bibfield  {title} {\bibinfo {title} {Magnetic-field-driven crack formation
  in an evaporated anisotropic colloidal assembly},\ }\href
  {https://doi.org/10.1103/PhysRevE.94.012618} {\bibfield  {journal} {\bibinfo
  {journal} {Phys. Rev. E}\ }\textbf {\bibinfo {volume} {94}},\ \bibinfo
  {pages} {012618} (\bibinfo {year} {2016})}\BibitemShut {NoStop}%
\bibitem [{\citenamefont {Szatmári}\ \emph {et~al.}(2021)\citenamefont
  {Szatmári}, \citenamefont {Halász}, \citenamefont {Nakahara}, \citenamefont
  {Kitsunezaki},\ and\ \citenamefont {Kun}}]{Szatmari2021}%
  \BibitemOpen
  \bibfield  {author} {\bibinfo {author} {\bibfnamefont {R.}~\bibnamefont
  {Szatmári}}, \bibinfo {author} {\bibfnamefont {Z.}~\bibnamefont {Halász}},
  \bibinfo {author} {\bibfnamefont {A.}~\bibnamefont {Nakahara}}, \bibinfo
  {author} {\bibfnamefont {S.}~\bibnamefont {Kitsunezaki}},\ and\ \bibinfo
  {author} {\bibfnamefont {F.}~\bibnamefont {Kun}},\ }\bibfield  {title}
  {\bibinfo {title} {Evolution of anisotropic crack patterns in shrinking
  material layers},\ }\href {https://doi.org/10.1039/D1SM01193F} {\bibfield
  {journal} {\bibinfo  {journal} {Soft Matter}\ }\textbf {\bibinfo {volume}
  {17}},\ \bibinfo {pages} {10005} (\bibinfo {year} {2021})}\BibitemShut
  {NoStop}%
\bibitem [{\citenamefont {Uemura}\ \emph {et~al.}(2024)\citenamefont {Uemura},
  \citenamefont {Nakahara}, \citenamefont {Matsuo},\ and\ \citenamefont
  {Iwata}}]{Uemura2024}%
  \BibitemOpen
  \bibfield  {author} {\bibinfo {author} {\bibfnamefont {C.}~\bibnamefont
  {Uemura}}, \bibinfo {author} {\bibfnamefont {A.}~\bibnamefont {Nakahara}},
  \bibinfo {author} {\bibfnamefont {Y.}~\bibnamefont {Matsuo}},\ and\ \bibinfo
  {author} {\bibfnamefont {T.}~\bibnamefont {Iwata}},\ }\bibfield  {title}
  {\bibinfo {title} {Transition condition between memories of vibration and
  flow in the memory effect of paste},\ }\href
  {https://doi.org/10.1103/PhysRevE.109.034604} {\bibfield  {journal} {\bibinfo
   {journal} {Phys. Rev. E}\ }\textbf {\bibinfo {volume} {109}},\ \bibinfo
  {pages} {034604} (\bibinfo {year} {2024})}\BibitemShut {NoStop}%
\bibitem [{\citenamefont {Baba}\ \emph {et~al.}(2023)\citenamefont {Baba},
  \citenamefont {Fujimaki}, \citenamefont {Uemura}, \citenamefont {Matsuo},
  \citenamefont {Nakahara},\ and\ \citenamefont {Muramatsu}}]{Baba2023}%
  \BibitemOpen
  \bibfield  {author} {\bibinfo {author} {\bibfnamefont {R.}~\bibnamefont
  {Baba}}, \bibinfo {author} {\bibfnamefont {K.}~\bibnamefont {Fujimaki}},
  \bibinfo {author} {\bibfnamefont {C.}~\bibnamefont {Uemura}}, \bibinfo
  {author} {\bibfnamefont {Y.}~\bibnamefont {Matsuo}}, \bibinfo {author}
  {\bibfnamefont {A.}~\bibnamefont {Nakahara}},\ and\ \bibinfo {author}
  {\bibfnamefont {A.}~\bibnamefont {Muramatsu}},\ }\bibfield  {title} {\bibinfo
  {title} {Assisting and eliminating memory effects of paste by adding
  polysaccharides},\ }\href {https://doi.org/10.1103/PhysRevE.108.054602}
  {\bibfield  {journal} {\bibinfo  {journal} {Phys. Rev. E}\ }\textbf {\bibinfo
  {volume} {108}},\ \bibinfo {pages} {054602} (\bibinfo {year}
  {2023})}\BibitemShut {NoStop}%
\bibitem [{\citenamefont {Nakahara}\ \emph {et~al.}(2011)\citenamefont
  {Nakahara}, \citenamefont {Shinohara},\ and\ \citenamefont
  {Matsuo}}]{Nakahara2011}%
  \BibitemOpen
  \bibfield  {author} {\bibinfo {author} {\bibfnamefont {A.}~\bibnamefont
  {Nakahara}}, \bibinfo {author} {\bibfnamefont {Y.}~\bibnamefont
  {Shinohara}},\ and\ \bibinfo {author} {\bibfnamefont {Y.}~\bibnamefont
  {Matsuo}},\ }\bibfield  {title} {\bibinfo {title} {Control of crack pattern
  using memory effect of paste},\ }\href
  {https://doi.org/10.1088/1742-6596/319/1/012014} {\bibfield  {journal}
  {\bibinfo  {journal} {J. Phys. Conf. Ser.}\ }\textbf {\bibinfo {volume}
  {319}},\ \bibinfo {pages} {012014} (\bibinfo {year} {2011})}\BibitemShut
  {NoStop}%
\bibitem [{\citenamefont {Nakahara}\ \emph {et~al.}(2019)\citenamefont
  {Nakahara}, \citenamefont {Hiraoka}, \citenamefont {Hayashi}, \citenamefont
  {Matsuo},\ and\ \citenamefont {Kitsunezaki}}]{Nakahara2019}%
  \BibitemOpen
  \bibfield  {author} {\bibinfo {author} {\bibfnamefont {A.}~\bibnamefont
  {Nakahara}}, \bibinfo {author} {\bibfnamefont {T.}~\bibnamefont {Hiraoka}},
  \bibinfo {author} {\bibfnamefont {R.}~\bibnamefont {Hayashi}}, \bibinfo
  {author} {\bibfnamefont {Y.}~\bibnamefont {Matsuo}},\ and\ \bibinfo {author}
  {\bibfnamefont {S.}~\bibnamefont {Kitsunezaki}},\ }\bibfield  {title}
  {\bibinfo {title} {Mechanism of memory effect of paste which dominates
  desiccation crack patterns},\ }\href {https://doi.org/10.1098/rsta.2017.0395}
  {\bibfield  {journal} {\bibinfo  {journal} {Philos. Trans. R. Soc. A}\
  }\textbf {\bibinfo {volume} {377}},\ \bibinfo {pages} {20170395} (\bibinfo
  {year} {2019})}\BibitemShut {NoStop}%
\bibitem [{\citenamefont {Kitsunezaki}(1999)}]{Kitsunezaki1999}%
  \BibitemOpen
  \bibfield  {author} {\bibinfo {author} {\bibfnamefont {S.}~\bibnamefont
  {Kitsunezaki}},\ }\bibfield  {title} {\bibinfo {title} {Fracture patterns
  induced by desiccation in a thin layer},\ }\href
  {https://doi.org/10.1103/PhysRevE.60.6449} {\bibfield  {journal} {\bibinfo
  {journal} {Phys. Rev. E}\ }\textbf {\bibinfo {volume} {60}},\ \bibinfo
  {pages} {6449} (\bibinfo {year} {1999})}\BibitemShut {NoStop}%
\bibitem [{\citenamefont {Singh}\ and\ \citenamefont
  {Tirumkudulu}(2007)}]{Shingh2007}%
  \BibitemOpen
  \bibfield  {author} {\bibinfo {author} {\bibfnamefont {K.~B.}\ \bibnamefont
  {Singh}}\ and\ \bibinfo {author} {\bibfnamefont {M.~S.}\ \bibnamefont
  {Tirumkudulu}},\ }\bibfield  {title} {\bibinfo {title} {Cracking in drying
  colloidal films},\ }\href {https://doi.org/10.1103/PhysRevLett.98.218302}
  {\bibfield  {journal} {\bibinfo  {journal} {Phys. Rev. Lett.}\ }\textbf
  {\bibinfo {volume} {98}},\ \bibinfo {pages} {218302} (\bibinfo {year}
  {2007})}\BibitemShut {NoStop}%
\bibitem [{\citenamefont {Man}\ and\ \citenamefont {Russel}(2008)}]{Man2008}%
  \BibitemOpen
  \bibfield  {author} {\bibinfo {author} {\bibfnamefont {W.}~\bibnamefont
  {Man}}\ and\ \bibinfo {author} {\bibfnamefont {W.~B.}\ \bibnamefont
  {Russel}},\ }\bibfield  {title} {\bibinfo {title} {Direct measurements of
  critical stresses and cracking in thin films of colloid dispersions},\ }\href
  {https://doi.org/10.1103/PhysRevLett.100.198302} {\bibfield  {journal}
  {\bibinfo  {journal} {Phys. Rev. Lett.}\ }\textbf {\bibinfo {volume} {100}},\
  \bibinfo {pages} {198302} (\bibinfo {year} {2008})}\BibitemShut {NoStop}%
\bibitem [{\citenamefont {Halász}\ \emph {et~al.}(2017)\citenamefont
  {Halász}, \citenamefont {Nakahara}, \citenamefont {Kitsunezaki},\ and\
  \citenamefont {Kun}}]{Halasz2017}%
  \BibitemOpen
  \bibfield  {author} {\bibinfo {author} {\bibfnamefont {Z.}~\bibnamefont
  {Halász}}, \bibinfo {author} {\bibfnamefont {A.}~\bibnamefont {Nakahara}},
  \bibinfo {author} {\bibfnamefont {S.}~\bibnamefont {Kitsunezaki}},\ and\
  \bibinfo {author} {\bibfnamefont {F.}~\bibnamefont {Kun}},\ }\bibfield
  {title} {\bibinfo {title} {Effect of disorder on shrinkage-induced
  fragmentation of a thin brittle layer},\ }\href
  {https://doi.org/10.1103/PhysRevE.96.033006} {\bibfield  {journal} {\bibinfo
  {journal} {Phys. Rev. E}\ }\textbf {\bibinfo {volume} {96}},\ \bibinfo
  {pages} {033006} (\bibinfo {year} {2017})}\BibitemShut {NoStop}%
\bibitem [{\citenamefont {Ito}\ and\ \citenamefont
  {Yukawa}(2014{\natexlab{a}})}]{Ito2014a}%
  \BibitemOpen
  \bibfield  {author} {\bibinfo {author} {\bibfnamefont {S.}~\bibnamefont
  {Ito}}\ and\ \bibinfo {author} {\bibfnamefont {S.}~\bibnamefont {Yukawa}},\
  }\bibfield  {title} {\bibinfo {title} {Dynamical scaling of fragment
  distribution in drying paste},\ }\href
  {https://doi.org/10.1103/PhysRevE.90.042909} {\bibfield  {journal} {\bibinfo
  {journal} {Phys. Rev. E}\ }\textbf {\bibinfo {volume} {90}},\ \bibinfo
  {pages} {042909} (\bibinfo {year} {2014}{\natexlab{a}})}\BibitemShut
  {NoStop}%
\bibitem [{\citenamefont {Ito}\ and\ \citenamefont
  {Yukawa}(2014{\natexlab{b}})}]{Ito2014b}%
  \BibitemOpen
  \bibfield  {author} {\bibinfo {author} {\bibfnamefont {S.}~\bibnamefont
  {Ito}}\ and\ \bibinfo {author} {\bibfnamefont {S.}~\bibnamefont {Yukawa}},\
  }\bibfield  {title} {\bibinfo {title} {Stochastic modeling on fragmentation
  process over lifetime and its dynamical scaling law of fragment
  distribution},\ }\href {https://doi.org/10.7566/JPSJ.83.124005} {\bibfield
  {journal} {\bibinfo  {journal} {J. Phys. Soc. Jpn.}\ }\textbf {\bibinfo
  {volume} {83}},\ \bibinfo {pages} {124005} (\bibinfo {year}
  {2014}{\natexlab{b}})}\BibitemShut {NoStop}%
\bibitem [{\citenamefont {Hirobe}\ and\ \citenamefont
  {Oguni}(2016)}]{Hirobe2016}%
  \BibitemOpen
  \bibfield  {author} {\bibinfo {author} {\bibfnamefont {S.}~\bibnamefont
  {Hirobe}}\ and\ \bibinfo {author} {\bibfnamefont {K.}~\bibnamefont {Oguni}},\
  }\bibfield  {title} {\bibinfo {title} {Coupling analysis of pattern formation
  in desiccation cracks},\ }\href {https://doi.org/10.1016/j.cma.2016.04.029}
  {\bibfield  {journal} {\bibinfo  {journal} {Comput. Methods Appl. Mech.
  Eng.}\ }\textbf {\bibinfo {volume} {307}},\ \bibinfo {pages} {470} (\bibinfo
  {year} {2016})}\BibitemShut {NoStop}%
\bibitem [{\citenamefont {Hirobe}(2017)}]{Hirobe2017a}%
  \BibitemOpen
  \bibfield  {author} {\bibinfo {author} {\bibfnamefont {S.}~\bibnamefont
  {Hirobe}},\ }\bibfield  {title} {\bibinfo {title} {Numerical simulation of
  desiccation cracking process by weak coupling of desiccation and fracture},\
  }\bibfield  {journal} {\bibinfo  {journal} {Int. J. GEOMATE}\ }\textbf
  {\bibinfo {volume} {12}},\ \href {https://doi.org/10.21660/2017.33.2535}
  {10.21660/2017.33.2535} (\bibinfo {year} {2017})\BibitemShut {NoStop}%
\bibitem [{\citenamefont {Hirobe}\ and\ \citenamefont
  {Oguni}(2017)}]{Hirobe2017b}%
  \BibitemOpen
  \bibfield  {author} {\bibinfo {author} {\bibfnamefont {S.}~\bibnamefont
  {Hirobe}}\ and\ \bibinfo {author} {\bibfnamefont {K.}~\bibnamefont {Oguni}},\
  }\bibfield  {title} {\bibinfo {title} {Modeling and numerical investigations
  for hierarchical pattern formation in desiccation cracking},\ }\href
  {https://doi.org/10.1016/j.physd.2017.08.002} {\bibfield  {journal} {\bibinfo
   {journal} {Physica D}\ }\textbf {\bibinfo {volume} {359}},\ \bibinfo {pages}
  {29} (\bibinfo {year} {2017})}\BibitemShut {NoStop}%
\bibitem [{\citenamefont {Hirobe}\ and\ \citenamefont
  {Oguni}(2019)}]{Hirobe2019}%
  \BibitemOpen
  \bibfield  {author} {\bibinfo {author} {\bibfnamefont {S.}~\bibnamefont
  {Hirobe}}\ and\ \bibinfo {author} {\bibfnamefont {K.}~\bibnamefont {Oguni}},\
  }\bibfield  {title} {\bibinfo {title} {Modeling and simulating methods for
  the desiccation cracking},\ }\href
  {https://doi.org/10.1142/S021987621840011X} {\bibfield  {journal} {\bibinfo
  {journal} {Int. J. Comput. Methods}\ }\textbf {\bibinfo {volume} {16}},\
  \bibinfo {pages} {1840011} (\bibinfo {year} {2019})}\BibitemShut {NoStop}%
\bibitem [{\citenamefont {Kitsunezaki}\ \emph {et~al.}(2016)\citenamefont
  {Kitsunezaki}, \citenamefont {Nakahara},\ and\ \citenamefont
  {Matsuo}}]{Kitsunezaki2016}%
  \BibitemOpen
  \bibfield  {author} {\bibinfo {author} {\bibfnamefont {S.}~\bibnamefont
  {Kitsunezaki}}, \bibinfo {author} {\bibfnamefont {A.}~\bibnamefont
  {Nakahara}},\ and\ \bibinfo {author} {\bibfnamefont {Y.}~\bibnamefont
  {Matsuo}},\ }\bibfield  {title} {\bibinfo {title} {Shaking-induced stress
  anisotropy in the memory effect of paste},\ }\href
  {https://doi.org/10.1209/0295-5075/114/64002} {\bibfield  {journal} {\bibinfo
   {journal} {EPL}\ }\textbf {\bibinfo {volume} {114}},\ \bibinfo {pages}
  {64002} (\bibinfo {year} {2016})}\BibitemShut {NoStop}%
\bibitem [{\citenamefont {Kitsunezaki}\ \emph {et~al.}(2017)\citenamefont
  {Kitsunezaki}, \citenamefont {Sasaki}, \citenamefont {Nishimoto},
  \citenamefont {Mizuguchi}, \citenamefont {Matsuo},\ and\ \citenamefont
  {Nakahara}}]{Kitsunezaki2017}%
  \BibitemOpen
  \bibfield  {author} {\bibinfo {author} {\bibfnamefont {S.}~\bibnamefont
  {Kitsunezaki}}, \bibinfo {author} {\bibfnamefont {A.}~\bibnamefont {Sasaki}},
  \bibinfo {author} {\bibfnamefont {A.}~\bibnamefont {Nishimoto}}, \bibinfo
  {author} {\bibfnamefont {T.}~\bibnamefont {Mizuguchi}}, \bibinfo {author}
  {\bibfnamefont {Y.}~\bibnamefont {Matsuo}},\ and\ \bibinfo {author}
  {\bibfnamefont {A.}~\bibnamefont {Nakahara}},\ }\bibfield  {title} {\bibinfo
  {title} {Memory effect and anisotropy of particle arrangements in granular
  paste},\ }\href {https://doi.org/10.1140/epje/i2017-11578-4} {\bibfield
  {journal} {\bibinfo  {journal} {Eur. Phys. J. E}\ }\textbf {\bibinfo {volume}
  {40}},\ \bibinfo {pages} {88} (\bibinfo {year} {2017})}\BibitemShut {NoStop}%
\bibitem [{\citenamefont {Otsuki}(2005)}]{Otsuki2005}%
  \BibitemOpen
  \bibfield  {author} {\bibinfo {author} {\bibfnamefont {M.}~\bibnamefont
  {Otsuki}},\ }\bibfield  {title} {\bibinfo {title} {Memory effect on the
  formation of drying cracks},\ }\bibfield  {journal} {\bibinfo  {journal}
  {Phys. Rev. E}\ }\textbf {\bibinfo {volume} {72}},\ \href
  {https://doi.org/10.1103/PhysRevE.72.046115} {10.1103/PhysRevE.72.046115}
  (\bibinfo {year} {2005})\BibitemShut {NoStop}%
\bibitem [{\citenamefont {Ooshida}(2008)}]{Ooshida2008}%
  \BibitemOpen
  \bibfield  {author} {\bibinfo {author} {\bibfnamefont {T.}~\bibnamefont
  {Ooshida}},\ }\bibfield  {title} {\bibinfo {title} {Continuum theory of
  memory effect in crack patterns of drying pastes},\ }\href
  {https://doi.org/10.1103/PhysRevE.77.061501} {\bibfield  {journal} {\bibinfo
  {journal} {Phys. Rev. E}\ }\textbf {\bibinfo {volume} {77}},\ \bibinfo
  {pages} {061501} (\bibinfo {year} {2008})}\BibitemShut {NoStop}%
\bibitem [{\citenamefont {Ooshida}(2009)}]{Ooshida2009}%
  \BibitemOpen
  \bibfield  {author} {\bibinfo {author} {\bibfnamefont {T.}~\bibnamefont
  {Ooshida}},\ }\bibfield  {title} {\bibinfo {title} {Three-dimensional
  residual tension theory of nakahara effect in pastes},\ }\href
  {https://doi.org/10.1143/JPSJ.78.104801} {\bibfield  {journal} {\bibinfo
  {journal} {J. Phys. Soc. Jpn.}\ }\textbf {\bibinfo {volume} {78}},\ \bibinfo
  {pages} {104801} (\bibinfo {year} {2009})}\BibitemShut {NoStop}%
\bibitem [{\citenamefont {Morita}\ and\ \citenamefont
  {Otsuki}(2021)}]{Morita2021}%
  \BibitemOpen
  \bibfield  {author} {\bibinfo {author} {\bibfnamefont {J.}~\bibnamefont
  {Morita}}\ and\ \bibinfo {author} {\bibfnamefont {M.}~\bibnamefont
  {Otsuki}},\ }\bibfield  {title} {\bibinfo {title} {Memory effect of external
  oscillation on residual stress in a paste},\ }\href
  {https://doi.org/10.1140/epje/s10189-021-00111-z} {\bibfield  {journal}
  {\bibinfo  {journal} {Eur. Phys. J. E}\ }\textbf {\bibinfo {volume} {44}},\
  \bibinfo {pages} {106} (\bibinfo {year} {2021})}\BibitemShut {NoStop}%
\bibitem [{\citenamefont {Marsden}\ and\ \citenamefont
  {Hughes}(1994)}]{Marsden1994}%
  \BibitemOpen
  \bibfield  {author} {\bibinfo {author} {\bibfnamefont {J.}~\bibnamefont
  {Marsden}}\ and\ \bibinfo {author} {\bibfnamefont {T.~J.~R.}\ \bibnamefont
  {Hughes}},\ }\href@noop {} {\emph {\bibinfo {title} {Mathematical foundations
  of elasticity}}}\ (\bibinfo  {publisher} {Dover Publications, Inc.},\
  \bibinfo {year} {1994})\BibitemShut {NoStop}%
\bibitem [{\citenamefont {Romano}\ and\ \citenamefont
  {Marasco}(2014)}]{Romano2014}%
  \BibitemOpen
  \bibfield  {author} {\bibinfo {author} {\bibfnamefont {A.}~\bibnamefont
  {Romano}}\ and\ \bibinfo {author} {\bibfnamefont {A.}~\bibnamefont
  {Marasco}},\ }\href {https://doi.org/10.1007/978-1-4939-1604-7} {\emph
  {\bibinfo {title} {Continuum Mechanics using Mathematica®}}}\ (\bibinfo
  {publisher} {Springer New York},\ \bibinfo {year} {2014})\BibitemShut
  {NoStop}%
\bibitem [{\citenamefont {Truesdell}\ and\ \citenamefont
  {Noll}(2004)}]{Truesdell2004}%
  \BibitemOpen
  \bibfield  {author} {\bibinfo {author} {\bibfnamefont {C.}~\bibnamefont
  {Truesdell}}\ and\ \bibinfo {author} {\bibfnamefont {W.}~\bibnamefont
  {Noll}},\ }\href {https://doi.org/10.1007/978-3-662-10388-3} {\emph {\bibinfo
  {title} {The Non-Linear Field Theories of Mechanics}}},\ edited by\ \bibinfo
  {editor} {\bibfnamefont {S.~S.}\ \bibnamefont {Antman}}\ (\bibinfo
  {publisher} {Springer Berlin Heidelberg},\ \bibinfo {year}
  {2004})\BibitemShut {NoStop}%
\bibitem [{\citenamefont {Beatty}(1987)}]{Beatty1987}%
  \BibitemOpen
  \bibfield  {author} {\bibinfo {author} {\bibfnamefont {M.~F.}\ \bibnamefont
  {Beatty}},\ }\bibfield  {title} {\bibinfo {title} {Topics in finite
  elasticity: Hyperelasticity of rubber, elastomers, and biological
  tissues—with examples},\ }\href {https://doi.org/10.1115/1.3149545}
  {\bibfield  {journal} {\bibinfo  {journal} {Appl. Mech. Rev.}\ }\textbf
  {\bibinfo {volume} {40}},\ \bibinfo {pages} {1699} (\bibinfo {year}
  {1987})}\BibitemShut {NoStop}%
\bibitem [{\citenamefont {Destrade}\ and\ \citenamefont
  {Saccomandi}(2005)}]{Destrade2005}%
  \BibitemOpen
  \bibfield  {author} {\bibinfo {author} {\bibfnamefont {M.}~\bibnamefont
  {Destrade}}\ and\ \bibinfo {author} {\bibfnamefont {G.}~\bibnamefont
  {Saccomandi}},\ }\bibfield  {title} {\bibinfo {title} {Finite amplitude
  elastic waves propagating in compressible solids},\ }\href
  {https://doi.org/10.1103/PhysRevE.72.016620} {\bibfield  {journal} {\bibinfo
  {journal} {Phys. Rev. E}\ }\textbf {\bibinfo {volume} {72}},\ \bibinfo
  {pages} {016620} (\bibinfo {year} {2005})}\BibitemShut {NoStop}%
\bibitem [{\citenamefont {Jones}(2009)}]{Jones2009}%
  \BibitemOpen
  \bibfield  {author} {\bibinfo {author} {\bibfnamefont {R.~M.}\ \bibnamefont
  {Jones}},\ }\href {https://books.google.co.jp/books?id=kiCVc3AJhVwC} {\emph
  {\bibinfo {title} {Deformation Theory of Plasticity}}}\ (\bibinfo
  {publisher} {Bull Ridge Pub.},\ \bibinfo {year} {2009})\BibitemShut {NoStop}%
\bibitem [{\citenamefont {Voigtmann}(2014)}]{Voigtmann2014}%
  \BibitemOpen
  \bibfield  {author} {\bibinfo {author} {\bibfnamefont {T.}~\bibnamefont
  {Voigtmann}},\ }\bibfield  {title} {\bibinfo {title} {Nonlinear glassy
  rheology},\ }\href {https://doi.org/10.1016/j.cocis.2014.11.001} {\bibfield
  {journal} {\bibinfo  {journal} {Curr. Opin. Colloid Interface Sci.}\ }\textbf
  {\bibinfo {volume} {19}},\ \bibinfo {pages} {549} (\bibinfo {year}
  {2014})}\BibitemShut {NoStop}%
\bibitem [{\citenamefont {Miyazaki}\ and\ \citenamefont
  {Reichman}(2002)}]{Miyazaki2002}%
  \BibitemOpen
  \bibfield  {author} {\bibinfo {author} {\bibfnamefont {K.}~\bibnamefont
  {Miyazaki}}\ and\ \bibinfo {author} {\bibfnamefont {D.~R.}\ \bibnamefont
  {Reichman}},\ }\bibfield  {title} {\bibinfo {title} {Molecular hydrodynamic
  theory of supercooled liquids and colloidal suspensions under shear},\ }\href
  {https://doi.org/10.1103/PhysRevE.66.050501} {\bibfield  {journal} {\bibinfo
  {journal} {Phys. Rev. E}\ }\textbf {\bibinfo {volume} {66}},\ \bibinfo
  {pages} {050501} (\bibinfo {year} {2002})}\BibitemShut {NoStop}%
\bibitem [{\citenamefont {Fuchs}\ and\ \citenamefont
  {Cates}(2002)}]{Fuchs2002}%
  \BibitemOpen
  \bibfield  {author} {\bibinfo {author} {\bibfnamefont {M.}~\bibnamefont
  {Fuchs}}\ and\ \bibinfo {author} {\bibfnamefont {M.~E.}\ \bibnamefont
  {Cates}},\ }\bibfield  {title} {\bibinfo {title} {Theory of nonlinear
  rheology and yielding of dense colloidal suspensions},\ }\href
  {https://doi.org/10.1103/PhysRevLett.89.248304} {\bibfield  {journal}
  {\bibinfo  {journal} {Phys. Rev. Lett.}\ }\textbf {\bibinfo {volume} {89}},\
  \bibinfo {pages} {248304} (\bibinfo {year} {2002})}\BibitemShut {NoStop}%
\bibitem [{\citenamefont {Ballauff}\ \emph {et~al.}(2013)\citenamefont
  {Ballauff}, \citenamefont {Brader}, \citenamefont {Egelhaaf}, \citenamefont
  {Fuchs}, \citenamefont {Horbach}, \citenamefont {Koumakis}, \citenamefont
  {Krüger}, \citenamefont {Laurati}, \citenamefont {Mutch}, \citenamefont
  {Petekidis}, \citenamefont {Siebenbürger}, \citenamefont {Voigtmann},\ and\
  \citenamefont {Zausch}}]{Ballauff2013}%
  \BibitemOpen
  \bibfield  {author} {\bibinfo {author} {\bibfnamefont {M.}~\bibnamefont
  {Ballauff}}, \bibinfo {author} {\bibfnamefont {J.~M.}\ \bibnamefont
  {Brader}}, \bibinfo {author} {\bibfnamefont {S.~U.}\ \bibnamefont
  {Egelhaaf}}, \bibinfo {author} {\bibfnamefont {M.}~\bibnamefont {Fuchs}},
  \bibinfo {author} {\bibfnamefont {J.}~\bibnamefont {Horbach}}, \bibinfo
  {author} {\bibfnamefont {N.}~\bibnamefont {Koumakis}}, \bibinfo {author}
  {\bibfnamefont {M.}~\bibnamefont {Krüger}}, \bibinfo {author} {\bibfnamefont
  {M.}~\bibnamefont {Laurati}}, \bibinfo {author} {\bibfnamefont {K.~J.}\
  \bibnamefont {Mutch}}, \bibinfo {author} {\bibfnamefont {G.}~\bibnamefont
  {Petekidis}}, \bibinfo {author} {\bibfnamefont {M.}~\bibnamefont
  {Siebenbürger}}, \bibinfo {author} {\bibfnamefont {T.}~\bibnamefont
  {Voigtmann}},\ and\ \bibinfo {author} {\bibfnamefont {J.}~\bibnamefont
  {Zausch}},\ }\bibfield  {title} {\bibinfo {title} {Residual stresses in
  glasses},\ }\href {https://doi.org/10.1103/PhysRevLett.110.215701} {\bibfield
   {journal} {\bibinfo  {journal} {Phys. Rev. Lett.}\ }\textbf {\bibinfo
  {volume} {110}},\ \bibinfo {pages} {215701} (\bibinfo {year}
  {2013})}\BibitemShut {NoStop}%
\bibitem [{\citenamefont {Mohan}\ \emph {et~al.}(2013)\citenamefont {Mohan},
  \citenamefont {Bonnecaze},\ and\ \citenamefont {Cloitre}}]{Mohan2013}%
  \BibitemOpen
  \bibfield  {author} {\bibinfo {author} {\bibfnamefont {L.}~\bibnamefont
  {Mohan}}, \bibinfo {author} {\bibfnamefont {R.~T.}\ \bibnamefont
  {Bonnecaze}},\ and\ \bibinfo {author} {\bibfnamefont {M.}~\bibnamefont
  {Cloitre}},\ }\bibfield  {title} {\bibinfo {title} {Microscopic origin of
  internal stresses in jammed soft particle suspensions},\ }\href
  {https://doi.org/10.1103/PhysRevLett.111.268301} {\bibfield  {journal}
  {\bibinfo  {journal} {Phys. Rev. Lett.}\ }\textbf {\bibinfo {volume} {111}},\
  \bibinfo {pages} {268301} (\bibinfo {year} {2013})}\BibitemShut {NoStop}%
\bibitem [{\citenamefont {Fritschi}\ \emph {et~al.}(2014)\citenamefont
  {Fritschi}, \citenamefont {Fuchs},\ and\ \citenamefont
  {Voigtmann}}]{Fritschi2014}%
  \BibitemOpen
  \bibfield  {author} {\bibinfo {author} {\bibfnamefont {S.}~\bibnamefont
  {Fritschi}}, \bibinfo {author} {\bibfnamefont {M.}~\bibnamefont {Fuchs}},\
  and\ \bibinfo {author} {\bibfnamefont {T.}~\bibnamefont {Voigtmann}},\
  }\bibfield  {title} {\bibinfo {title} {Mode-coupling analysis of residual
  stresses in colloidal glasses},\ }\href {https://doi.org/10.1039/C4SM00247D}
  {\bibfield  {journal} {\bibinfo  {journal} {Soft Matter}\ }\textbf {\bibinfo
  {volume} {10}},\ \bibinfo {pages} {4822} (\bibinfo {year}
  {2014})}\BibitemShut {NoStop}%
\bibitem [{\citenamefont {Mohan}\ \emph {et~al.}(2015)\citenamefont {Mohan},
  \citenamefont {Cloitre},\ and\ \citenamefont {Bonnecaze}}]{Mohan2015}%
  \BibitemOpen
  \bibfield  {author} {\bibinfo {author} {\bibfnamefont {L.}~\bibnamefont
  {Mohan}}, \bibinfo {author} {\bibfnamefont {M.}~\bibnamefont {Cloitre}},\
  and\ \bibinfo {author} {\bibfnamefont {R.~T.}\ \bibnamefont {Bonnecaze}},\
  }\bibfield  {title} {\bibinfo {title} {Build-up and two-step relaxation of
  internal stress in jammed suspensions},\ }\href
  {https://doi.org/10.1122/1.4901750} {\bibfield  {journal} {\bibinfo
  {journal} {J. Rheol.}\ }\textbf {\bibinfo {volume} {59}},\ \bibinfo {pages}
  {63} (\bibinfo {year} {2015})}\BibitemShut {NoStop}%
\bibitem [{\citenamefont {Moghimi}\ \emph {et~al.}(2017)\citenamefont
  {Moghimi}, \citenamefont {Jacob},\ and\ \citenamefont
  {Petekidis}}]{Moghimi2017}%
  \BibitemOpen
  \bibfield  {author} {\bibinfo {author} {\bibfnamefont {E.}~\bibnamefont
  {Moghimi}}, \bibinfo {author} {\bibfnamefont {A.~R.}\ \bibnamefont {Jacob}},\
  and\ \bibinfo {author} {\bibfnamefont {G.}~\bibnamefont {Petekidis}},\
  }\bibfield  {title} {\bibinfo {title} {Residual stresses in colloidal gels},\
  }\href {https://doi.org/10.1039/C7SM01655G} {\bibfield  {journal} {\bibinfo
  {journal} {Soft Matter}\ }\textbf {\bibinfo {volume} {13}},\ \bibinfo {pages}
  {7824} (\bibinfo {year} {2017})}\BibitemShut {NoStop}%
\bibitem [{\citenamefont {Otsuki}\ and\ \citenamefont
  {Sasa}(2006)}]{Otsuki2006}%
  \BibitemOpen
  \bibfield  {author} {\bibinfo {author} {\bibfnamefont {M.}~\bibnamefont
  {Otsuki}}\ and\ \bibinfo {author} {\bibfnamefont {S.}~\bibnamefont {Sasa}},\
  }\bibfield  {title} {\bibinfo {title} {An order parameter equation for the
  dynamic yield stress in dense colloidal suspensions},\ }\href
  {https://doi.org/10.1088/1742-5468/2006/10/L10004} {\bibfield  {journal}
  {\bibinfo  {journal} {J. Stat. Mech. Theor. Exp.}\ }\textbf {\bibinfo
  {volume} {2006}},\ \bibinfo {pages} {L10004} (\bibinfo {year}
  {2006})}\BibitemShut {NoStop}%
\end{thebibliography}%
	\end{document}